\documentclass{article}

\usepackage[utf8]{inputenc}
\usepackage{amsmath,amsfonts,amsthm} 
\usepackage{mathtools}
\usepackage[left=1.9cm, right=1.9cm, top=1.2cm, bottom=1.8cm,]{geometry}
\usepackage[numbers]{natbib}
\usepackage{float}
\usepackage{tikz}

\usepackage{caption}
\usepackage{subcaption}

\usepackage{authblk}

\usepackage{graphicx}
\graphicspath{ {./img/} }

\usepackage{xcolor}
\DeclareRobustCommand{\legendsquare}[1]{
  \textcolor{#1}{\rule{3ex}{2ex}}
}
\definecolor{yellow_mb}{rgb}{1, 1, 0}
\definecolor{yellow_red_mb_1}{rgb}{1, 0.667, 0}
\definecolor{yellow_red_mb_2}{rgb}{1, 0.333, 0}
\definecolor{red_mb}{rgb}{1, 0, 0}

\usepackage[title]{appendix}

\providecommand{\keywords}[1]
{
  \small  
  \textbf{\textit{Keywords---}} #1
}

\title{Bayesian I-optimal designs for choice experiments with mixtures}
\date{\vspace{-5ex}} 

\author{Mario Becerra$^1$}
\author{Peter Goos$^{1,2}$}
\affil{$^1$ Faculty of Bioscience Engineering, KU Leuven, Leuven, Belgium}
\affil{$^2$ Faculty of Business and Economics, Universiteit Antwerpen, Antwerpen, Belgium}

\newcommand{\set}[1]{\left\{ #1 \right\}}

\begin{document}

\maketitle

\begin{abstract}
\noindent Discrete choice experiments are frequently used to quantify consumer preferences by having respondents choose between different alternatives. Choice experiments involving mixtures of ingredients have been largely overlooked in the literature, even though many products and services can be described as mixtures of ingredients. As a consequence, little research has been done on the optimal design of choice experiments involving mixtures. The only existing research has focused on D-optimal designs, which means that an estimation-based approach was adopted. However, in experiments with mixtures, it is crucial to obtain models that yield precise predictions for any combination of ingredient proportions. This is because the goal of mixture experiments generally is to find the mixture that optimizes the respondents' utility. As a result, the I-optimality criterion is more suitable for designing choice experiments with mixtures than the D-optimality criterion because the I-optimality criterion focuses on getting precise predictions with the estimated statistical model. In this paper, we study Bayesian I-optimal designs, compare them with their Bayesian D-optimal counterparts, and show that the former designs perform substantially better than the latter in terms of the variance of the predicted utility.
\end{abstract}


\section{Introduction}

Discrete choice experiments are frequently used to quantify consumer preferences. These experiments collect stated preference data and are carried out by presenting respondents sets of alternatives, called choice sets, and having them choose between the alternatives. The respondents repeat this task several times with different choice sets. It is common to have the respondents do pairwise comparisons with choice sets that involve two alternatives each.

Discrete choice experiments have been successfully applied in areas such as marketing \cite{rossi2012bayesian, train_discrete_2009}, transportation \cite{zijlstra2019mixture}, health care \cite{luyten2015public}, ecology \cite{fletcher2015affinity, melero2018ecological, vardakis2015discrete}, and environmental economics \cite{bennett2001choice, torres2013payments, vojavcek2010comparison}. However, choice experiments involving mixtures have been largely overlooked in the literature, even though many products and services can be described as mixtures of ingredients.

Examples of mixture ingredients include the chemicals that are used to create a pesticide \cite{cornell_primer_2011}; media used in advertising campaigns \cite{goos2019using}; components of a mobility budget such as car with fuel card and public transport card \cite{zijlstra2019mixture}; cement, water, and sand to make concrete \cite{cornell2002experiments}; the wheat varieties used to bake bread \cite{rehman2007optimisation}; and the ingredients used to make a drink such as mango juice, lime juice, and blackcurrant syrup \cite{courcoux1997methode, goos_hamidouche_2019_choice}. In mixture experiments, the products and services under investigation are expressed as combinations of proportions of ingredients and the researchers' interest is generally in one or more characteristics of the mixture. In this paper, the characteristic of interest is the preference of respondents. Choice experiments are ideal to collect data for quantifying and modeling preferences for mixtures.

The first example of a discrete choice experiment concerning mixtures was published by \citeauthor{courcoux1997methode} \cite{courcoux1997methode}. The experiment is actually a taste experiment to model the preferences for cocktails involving different proportions of mango juice, lime juice, and blackcurrant syrup. The resulting experimental data involve the responses of sixty people, each making eight pairwise comparisons of different cocktails. Thus, eight times in total,  each respondent had to taste two different cocktails and say which one they preferred. This means that the choice experiment comprised $60 \times 8 = 480$ choice sets, each of size 2. \citeauthor{goos_hamidouche_2019_choice} \cite{goos_hamidouche_2019_choice} described how to combine Scheffé models for data from mixture experiments with the logit type models typically used for choice experiments, and demonstrated the usefulness of the resulting combined models using the data from \citeauthor{courcoux1997methode} \cite{courcoux1997methode}.

As discrete choice experiments in general, and taste experiments in the form of a discrete choice experiment in particular, are expensive, cumbersome and time-consuming, efficient experimental designs are required so that the experiments provide reliable information for statistical modeling, precise estimation of model parameters and precise predictions. Optimal design of experiments is the branch of statistics dealing with the construction of efficient experimental designs. 

Little research has been done concerning the optimal design of choice experiments with mixtures. For their experiment, \citeauthor{courcoux1997methode} \cite{courcoux1997methode} used an elegant ad hoc design construction combining a simplex centroid design to define the mixtures used and a balanced incomplete block design to define subsets of the mixtures. Assuming a multinomial logit model, \citeauthor{ruseckaite_bayesian_2017} \cite{ruseckaite_bayesian_2017} compared two algorithms to find D-optimal designs for choice experiments with mixtures. A D-optimal design is an experimental design that maximizes the determinant of the Fisher information matrix. The D-optimal design approach can be viewed as an estimation-based approach, because it is intended to minimize the generalized variance of the estimators of the model parameters. D-optimal experimental designs are thus a good choice if obtaining low-variance estimators is the main goal of the choice experiment. However, in experiments with mixtures, the goal generally is to optimize the composition of the mixture to maximize consumer preference. Therefore, in mixture experiments, it is crucial to obtain models that yield precise predictions for any combination of ingredient proportions. As a result, I-optimal experimental designs are more suitable for choice experiments with mixtures than D-optimal ones. I-optimal designs minimize the average prediction variance and therefore allow a better identification of the mixture that maximizes the consumer preference. In this paper, we therefore study I-optimal designs for discrete choice experiments involving mixtures.

The rest of the paper is organized as follows. In Section \ref{sec:models}, we describe how to incorporate mixtures into the multinomial logit choice model. In Section \ref{sec:opt_des_criteria}, we describe the D- and I-optimality criteria. In Section \ref{sec:construction_of_iopt_designs}, we outline the design construction algorithm we use. In Section \ref{sec:results}, we present our computational results. Finally, in Section \ref{sec:discussion}, we summarize and discuss our work and include some suggestions for future research.


\section{Models}\label{sec:models}

In this section, we introduce the most commonly used models for data from mixture experiments as well as the multinomial logit model for choice data, and explain how to combine the two models for data from choice experiments involving mixtures.

\subsection{Models for data from mixture experiments}
\label{sectionwithconstraints}

Mixture experiments involve two or more ingredients and a response variable that depends only on the relative proportions of the ingredients in the mixture. Each mixture is described as a combination of $q$ ingredient proportions, with the constraint that these proportions sum up to one. Due to this constraint, a classical regression model involving an intercept and linear terms in the ingredient proportions exhibits perfect collinearity. Therefore, researchers must use dedicated regression models when analyzing data from mixture experiments. The most commonly used family of models for data from mixture experiments is the Scheffé family \cite{scheffe1958experiments, scheffe1963simplex}. The most popular Scheffé models are the first-order, second-order, and special-cubic models. 

Denoting the response in a traditional mixture experiment with a continuous outcome by $Y$ and the $q$ ingredient proportions by $x_1,x_2,\dots,x_q$, with $x_i \geq 0$ and $\sum_{i = 1}^q x_i = 1$, the first-order Scheffé model is
\begin{equation*}\label{eq_scheffe_first_order}
    Y = \sum_{i = 1}^q \beta_i x_i + \varepsilon.
\end{equation*}
The second-order Scheffé model is
\begin{equation*}\label{eq_scheffe_second_order}
    Y = 
      \sum_{i = 1}^q \beta_i x_i + 
      \sum_{i = 1}^{q-1} \sum_{j = i+1}^q \beta_{ij} x_i x_j + 
      \varepsilon,
\end{equation*}
and, finally, the special-cubic Scheffé model is
\begin{equation*}\label{eq_scheffe_special_cubic}
    Y = 
      \sum_{i = 1}^q \beta_i x_i + 
      \sum_{i = 1}^{q-1} \sum_{j = i+1}^q \beta_{ij} x_i x_j + 
      \sum_{i = 1}^{q-2} \sum_{j = i+1}^{q-1} \sum_{k = j+1}^{q} \beta_{ijk} x_i x_j x_k + 
      \varepsilon.
\end{equation*}
In all three cases, $\varepsilon$ denotes the error term, which, traditionally, is assumed to be normally distributed.

\subsection{Multinomial logit model for choice data}

The multinomial logit model builds on random-utility theory and assumes that a respondent in a choice experiment faces $S$ choice sets involving $J$ alternatives. The model assumes that, within each choice set $s \in \set{1, ..., S}$, each respondent chooses the alternative that has the highest perceived utility. Therefore, the probability that a respondent chooses alternative $j \in \set{1, ..., J}$ in choice set $s$, denoted by $p_{js}$, is the probability that the perceived utility of alternative $j$ in choice set $s$, denoted by $U_{js}$, is larger than that of the other alternatives in the choice set:
\begin{equation*}
    \label{eq_p_js_max}
    p_{js} = \mathbb{P} \left[ U_{js} > \max(U_{1s}, ..., U_{j-1, s}, U_{j+1, s}, ..., U_{Js} ) \right].
\end{equation*}
Since each alternative in a choice set has a set of observable attributes that characterize it, the perceived utility $U_{js}$ can be expressed as 
\begin{equation}
    \label{eq_utility_js}
    U_{js} = \boldsymbol{f}^T(\boldsymbol{x}_{js}) \boldsymbol{\beta} + \varepsilon_{js},
\end{equation}
where $\boldsymbol{x}_{js}$ is the vector that contains the attributes corresponding to alternative $j$ in choice set $s$, $\boldsymbol{f}(\boldsymbol{x}_{js})$ represents the model expansion of this attribute vector, and $\boldsymbol{\beta}$ is the vector containing the model parameters. The model parameters contained within $\boldsymbol{\beta}$ express the preferences of the respondents for the alternatives' attributes. The error terms $\varepsilon_{js}$ in Equation~(\ref{eq_utility_js}) are assumed to be independent and identically Gumbel distributed. The Gumbel distribution is also known as the generalized extreme value distribution type I and as the log-Weibull distribution. As a result of the distributional assumption, it can be shown that
\begin{equation*}
    \label{eq_p_j}
    p_{js} = \frac{ \exp{ \left[ \boldsymbol{f}^T(\boldsymbol{x}_{js}) \boldsymbol{\beta} \right]} }{ \sum_{t = 1}^J \exp{ \left[ \boldsymbol{f}^T(\boldsymbol{x}_{ts}) \boldsymbol{\beta} \right]} },
\end{equation*}
which is called the softmax function in some research areas.

\subsection{Model for choice data concerning mixtures}\label{subsec:mixtures_and_mnl_model}

In this paper, where we focus on choice experiments involving mixtures, we assume that the attributes of the alternatives in a choice experiment are the ingredients of a mixture. Consequently, we assume that the attribute vector $\boldsymbol{x}_{js}$ contains the $q$ ingredient proportions $x_1,x_2,\dots,x_q$ and that $\boldsymbol{f}(\boldsymbol{x}_{js})$ represents the model expansion of these proportions. Following \citeauthor{ruseckaite_bayesian_2017} \cite{ruseckaite_bayesian_2017}, we base the polynomial expansion $\boldsymbol{f}(\boldsymbol{x}_{js})$ on the first-order, second-order, or special-cubic Scheffé model; depending on the complexity of the respondents' preferences. To explain how this is done, we start from the special-cubic model. The derivation of the models based on the first-order and second-order Scheffé model is analogous.

When starting from the special-cubic Scheffé model, the most natural thing to do would be to write the perceived utility $U_{js}$ of a mixture alternative $j$ in choice set $s$ as
\begin{equation*}
    \label{eq_u_js_mixt_0}
    U_{js} = 
    \sum_{i = 1}^{q} \beta_i x_{ijs} + 
    \sum_{i = 1}^{q-1} \sum_{k = i + 1}^{q} \beta_{ik} x_{ijs} x_{kjs} + 
    \sum_{i = 1}^{q-2} \sum_{k = i + 1}^{q-1} \sum_{l = k + 1}^{q} \beta_{ikl} x_{ijs} x_{kjs} x_{ljs} + \varepsilon_{js},
\end{equation*}
with the error terms $\varepsilon_{js}$ assumed to be independent and identically Gumbel distributed. However, due to the constraint that the ingredient proportions must sum up to one, this leads to an inestimable multinomial logit model. As a matter of fact, due to that constraint, we can rewrite $x_{qjs}$ as $1 - x_{1js} - ... - x_{q-1,js}$ and $U_{js}$ as
\begin{equation*}
    \label{eq_u_js_mixt_unidetif}
    \begin{aligned}
    U_{js} 
    &= 
    \sum_{i = 1}^{q-1} \beta_i x_{ijs} + \beta_q (1 - x_{1js} - ... - x_{q-1,j,s}) + 
    \sum_{i = 1}^{q-1} \sum_{k = i + 1}^{q} \beta_{ik} x_{ijs} x_{kjs} + 
    \sum_{i = 1}^{q-2} \sum_{k = i + 1}^{q-1} \sum_{l = k + 1}^{q} \beta_{ikl} x_{ijs} x_{kjs} x_{ljs} + \varepsilon_{js} \\ 
    &=
    \beta_q + \sum_{i = 1}^{q-1} (\beta_i - \beta_q) x_{ijs} + 
    \sum_{i = 1}^{q-1} \sum_{k = i + 1}^{q} \beta_{ik} x_{ijs} x_{kjs} + 
    \sum_{i = 1}^{q-2} \sum_{k = i + 1}^{q-1} \sum_{l = k + 1}^{q} \beta_{ikl} x_{ijs} x_{kjs} x_{ljs} + \varepsilon_{js}.
    \end{aligned}
\end{equation*}
This final expression for the perceived utility $U_{js}$ involves a constant, $\beta_q$. Because the multinomial logit model only takes into account differences in utility, that constant causes any choice model based on $U_{js}$ to be ill-defined and therefore inestimable. This can be easily remedied by dropping $\beta_q$ and using the following expression for the perceived utility:
\begin{equation*}
    U_{js} = 
    \boldsymbol{f}^T(\boldsymbol{x}_{js}) \boldsymbol{\beta} + \varepsilon_{js} =
    \sum_{i = 1}^{q-1} \beta_i^* x_{ijs} + 
    \sum_{i = 1}^{q-1} \sum_{k = i + 1}^{q} \beta_{ik} x_{ijs} x_{kjs} + 
    \sum_{i = 1}^{q-2} \sum_{k = i + 1}^{q-1} \sum_{l = k + 1}^{q} \beta_{ikl} x_{ijs} x_{kjs} x_{ljs} +
    \varepsilon_{js},
\end{equation*}
with $\beta_i^* = \beta_i - \beta_q$ for $i \in \set{1, ..., q-1}$. In this expression,
\[
\boldsymbol{x}_{js} = \left( x_{1js},x_{2js},\dots,x_{qjs} \right)^T
\]
and 
\[
\boldsymbol{f}(\boldsymbol{x}_{js}) =
\left( 
x_{1js},x_{2js},\dots,x_{q-1,js},
x_{1js}x_{2js},\dots,x_{q-1,js}x_{qjs},
x_{1js}x_{2js}x_{3js},\dots,x_{q-2,js}x_{q-1,js}x_{qjs}
\right)^T.
\]
The parameter vector $\boldsymbol{\beta}$ for the special-cubic model is then given by
$$
    \boldsymbol{\beta} = 
    \left(
    \beta_{1}^*, \beta_{2}^*, ..., \beta_{q-1}^*, \beta_{1,2}, ..., \beta_{q-1,q}, \beta_{123}, ..., \beta_{q-2,q-1,q}
    \right)^T.
$$
This vector has $r = (q^3 + 5q)/6-1$ elements in the event the special-cubic model is used. For the first- and second-order Scheffé models, the parameter vector $\boldsymbol{\beta}$ involves $r = q-1$ and $r = (q^2 + q)/2-1$ elements, respectively, after implementing the remedy to make the model estimable.


\section{Optimal design criteria}
\label{sec:opt_des_criteria}

\subsection{Information matrix}

To select D- and I-optimal experimental designs, it is required to know the information matrix of the model. For the multinomial logit model, the information matrix $\boldsymbol{I}(\boldsymbol{X}, \boldsymbol{\beta})$ is obtained as the sum of the information matrices of each of the $S$ choice sets \cite{kessels_comparison_2006}:
\begin{equation*}
    \label{eq_inf_matrix}
    \boldsymbol{I}(\boldsymbol{X}, \boldsymbol{\beta}) = 
        \sum_{s = 1}^S 
        \boldsymbol{X}_s^T (\boldsymbol{P}_s - \boldsymbol{p}_s \boldsymbol{p}_s^T) \boldsymbol{X}_s,
\end{equation*}
with $\boldsymbol{p}_s = \left( p_{1s}, ..., p_{Js} \right)^T$, $\boldsymbol{P}_s = \mathrm{diag}(\boldsymbol{p}_s)$, $\boldsymbol{X}_s^T = \left[ \boldsymbol{f}(\boldsymbol{x}_{js}) \right]_{j \in \set{ 1, ..., J }}$ denoting the model matrix corresponding to all alternatives in choice set $s$, and $\boldsymbol{X} = \left[ \boldsymbol{X}_1, ..., \boldsymbol{X}_S \right]$ denoting the model matrix for all choice sets. The information matrix $\boldsymbol{I}(\boldsymbol{X}, \boldsymbol{\beta})$ is of dimension $r \times r$. The inverse of the information matrix is the asymptotic variance-covariance matrix of the parameter estimates.

Note that the information matrix depends on the unknown parameter vector $\boldsymbol{\beta}$, through the choice probabilities contained within $\boldsymbol{p}_s$ and $\boldsymbol{P}_s$. This is typical for models that are not linear in the parameters, such as discrete choice models, and it implies that prior information is needed to find optimal designs \cite{atkinson199614, kessels_comparison_2006, ruseckaite_bayesian_2017}. The prior information can be in the form of a point estimate, or in the form of a prior distribution. An optimal design that uses only a point estimate is called a locally optimal design, whilst one that uses a prior distribution is called a Bayesian optimal design. 

In optimal experimental design, there are different criteria to measure the quality of a design and the corresponding model matrix $\boldsymbol{X}$. As pointed out in \citeauthor{goos_jones_optimal_2011} \cite{goos_jones_optimal_2011}, the two most widely used criteria for selecting experimental designs in business and industry are D-optimality and I-optimality (also called V-optimality; see \citeauthor{gosy} \cite{gosy}). The former is an estimation-oriented criterion because it focuses on a precise model estimation, while the latter is a prediction-oriented criterion because it focuses on getting precise predictions with the estimated statistical model.

\subsection{D-optimal designs}

The D-optimality criterion is the most traditional metric used in the design of choice experiments \cite{bliemer2009efficient, bliemer2010construction, bliemer2011experimental, burgess2005optimal, grasshoff2003optimal, kessels_usefulness_2011}. For a model matrix $\boldsymbol{X}$ and parameter vector $\boldsymbol{\beta}$, the D-optimality criterion can be defined as
\begin{equation}\label{eq_d_optim_classical}
    \mathcal{D} =
    \log{\left( 
    \det \left( 
         \left[ 
            \boldsymbol{I}^{-1}(\boldsymbol{X}, \boldsymbol{\beta})
         \right) \right]^{\frac{1}{r}}
    \right)},
\end{equation}
where $\boldsymbol{I}^{-1}(\boldsymbol{X}, \boldsymbol{\beta})$ is the asymptotic variance-covariance matrix and the inverse of the information matrix of the parameter estimates. A design that minimizes Equation~(\ref{eq_d_optim_classical}) using a point estimate of $\boldsymbol{\beta}$ is called a locally D-optimal design. The problem with locally D-optimal designs is that they may perform poorly for values of the parameter vector $\boldsymbol{\beta}$ for which they were not optimized. This weakness is, of course, highly relevant because the true values of the model parameters are unknown.

Bayesian designs take into account prior information and uncertainty about the parameter vector $\boldsymbol{\beta}$. More specifically, they are based on a prior distribution $\pi(\boldsymbol{\beta})$ which summarizes the prior knowledge concerning $\boldsymbol{\beta}$. Most of the Bayesian design constructions in the choice experiments literature adopt the approach where the D-optimality criterion is averaged over the prior distribution \cite{bliemer2009efficient, bliemer2011experimental, kessels_usefulness_2011}.
This is also the approach we use. Therefore, following \citeauthor{ruseckaite_bayesian_2017} \cite{ruseckaite_bayesian_2017}, we define the Bayesian D-optimality criterion for the multinomial logit model as
\begin{equation}
    \label{eq_pseudo_bayes_d_eff}
    \mathcal{D}_B =
    \log{\left( 
        \int_{\mathbb{R}^{r}}
        \left[  
            \det \left( 
                \boldsymbol{I}^{-1}(\boldsymbol{X}, \boldsymbol{\beta})
             \right) \right]^{\frac{1}{r}} 
        \pi(\boldsymbol{\beta}) d\boldsymbol{\beta}
        \right)},
\end{equation}
\noindent
where $\pi(\boldsymbol{\beta})$ is the prior distribution of $\boldsymbol{\beta}$. We refer to a design that minimizes the Bayesian D-optimality criterion as a Bayesian D-optimal design, even though the criterion does not take into account the posterior distribution and some authors therefore prefer to call these designs pseudo-Bayesian designs \cite{ryan2016review}.

\subsection{I-optimal designs}

The I-optimality criterion is generally defined as the average of the prediction variance over the experimental region, which we denote by $\chi$ and which is the $(q-1)$-dimensional simplex if there are no constraints on the ingredient proportions other than those mentioned in Section~\ref{sectionwithconstraints}. Now, when using choice models, there are two ways in which we can define I-optimality. If the goal is to predict choice probabilities, the I-optimality criterion is the average variance of the predicted choice probabilities. If the goal is to predict perceived utilities, the I-optimality criterion is the average variance of the predicted utilities. 

\subsubsection{I-optimality for predicted choice probabilities} 

\citeauthor{kessels_efficient_2009} \cite{kessels_efficient_2009} computed I-optimal designs based on the variance of the predicted choice probability, $\mathrm{Var} \left[ \hat{p}_{js} \right]$. Since this prediction variance cannot be calculated analytically, they approximated it using a first-order Taylor series expansion of the choice probability: 
\begin{equation*}
    \mathrm{Var} \left[ \hat{p}_{js} \right] \approx \boldsymbol{c}^T(\boldsymbol{x}_{js}) \boldsymbol{I}^{-1}(\boldsymbol{X}, \boldsymbol{\beta}) \boldsymbol{c}(\boldsymbol{x}_{js}),
\end{equation*}
where $\hat{p}_{js}$ denotes the predicted choice probability, and
\begin{equation*}
    \label{eq_derivative_p_beta}
    \boldsymbol{c}(\boldsymbol{x}_{js}) =
    \frac{\partial p_{js}}{\partial \boldsymbol{\beta}} =
    p_{js} \left( \boldsymbol{x}_{js} - \sum_{t = 1}^J p_{ts} \boldsymbol{x}_{ts} \right).
\end{equation*}
As a consequence, the I-optimality criterion of \citeauthor{kessels_efficient_2009} \cite{kessels_efficient_2009} is
\begin{equation}
    \label{eq_i_optim_classical_1}
    \frac{\int_{\chi} \mathrm{Var} \left[ \hat{p}_{js} \right] d\boldsymbol{x}_{js}}{\int_{\chi} d\boldsymbol{x}_{js}}
    =
     \frac{\int_{\chi} \boldsymbol{c}^T(\boldsymbol{x}_{js}) \boldsymbol{I}^{-1}(\boldsymbol{X}, \boldsymbol{\beta}) \boldsymbol{c}(\boldsymbol{x}_{js}) d\boldsymbol{x}_{js}}{\int_{\chi} d\boldsymbol{x}_{js}}
  =
    \frac{\int_{\chi} \mathrm{tr} \left[ \boldsymbol{I}^{-1}(\boldsymbol{X}, \boldsymbol{\beta}) \boldsymbol{c}(\boldsymbol{x}_{js}) \boldsymbol{c}^T(\boldsymbol{x}_{js}) \right] d\boldsymbol{x}_{js}}{\int_{\chi} d\boldsymbol{x}_{js}}
     =
     \frac{\mathrm{tr}\left[  \boldsymbol{I}^{-1}(\boldsymbol{X}, \boldsymbol{\beta}) \boldsymbol{W}_p(\boldsymbol{\beta}) \right]}{\int_{\chi} d\boldsymbol{x}_{js}},
\end{equation}
where 
\begin{equation}
    \label{eq_moments_matrix_orig}
    \boldsymbol{W}_p(\boldsymbol{\beta}) = \int_{\chi} \boldsymbol{c}(\boldsymbol{x}_{js}) \boldsymbol{c}^T(\boldsymbol{x}_{js}) d\boldsymbol{x}_{js}
\end{equation}
is referred to as the moments matrix in the literature on I-optimality and the subscript $p$ refers to the fact that this is the moments matrix for the I-optimality criterion based on choice probabilities. The denominator in Equation~(\ref{eq_i_optim_classical_1}) is the volume of the experimental region $\chi$. This denominator can be safely ignored when constructing I-optimal designs because it is constant for all designs for a given experiment.

Due to the fact that \citeauthor{kessels_efficient_2009} \cite{kessels_efficient_2009} considered only categorical attributes in their choice experiments, they were able to calculate the moments matrix exactly for any given parameter vector $\boldsymbol{\beta}$. This is impossible when continuous attributes are considered, because there is no closed-form solution for the integral in Equation (\ref{eq_moments_matrix_orig}). Since the attributes we consider in this paper are ingredient proportions, we also cannot calculate $\boldsymbol{W}_p(\boldsymbol{\beta})$ exactly. The solution to this problem would be to numerically approximate the integral in Equation (\ref{eq_moments_matrix_orig}). 

In our view, basing the I-optimality criterion for choice experiments with mixtures on predicted probabilities suffers from four weaknesses. First, there is no analytical expression for the variance of a predicted choice probability, which necessitates a first approximation. Second, there is no closed form expression for the average of the approximation of the variance of the predicted choice probability, which necessitates a second approximation. Third, the fact that $\boldsymbol{W}_p(\boldsymbol{\beta})$ depends on $\boldsymbol{\beta}$ implies that it increases the computational burden in the event a Bayesian optimal design is desired. Finally, any choice probability is always calculated with respect to a certain choice set, involving a certain number of alternatives. Hence, to be able to use the I-optimality criterion based on predicted probabilities, we need to pick a choice set size, a number of choice sets to consider and alternatives to be included in these choice sets. As there is no reason to prefer one choice set size, one number of choice sets, or one set of alternatives over another; there is no indisputable way to define the I-optimality criterion based on choice probabilities. Due to these weaknesses, we prefer to define the I-optimality criterion for choice experiments with mixtures based on predicted utilities.

\subsubsection{I-optimality for predicted utilities} 

In this paper, we base the I-optimality criterion on predicted utilities. One reason to do so is that that approach does not suffer from the four weaknesses we identified for the I-optimality criterion based on predicted choice probabilities. Another reason for our approach is that, to find the mixture that maximizes the consumer preferences, it suffices to identify the mixture that maximizes the predicted utility. As a matter of fact, the mixture with the largest possible predicted utility will automatically have the largest predicted choice probability, regardless of the alternatives to it. Finally, starting from predicted utilities is mathematically elegant because there is a closed form expression for the variance of the predicted perceived utility, for a given parameter vector $\boldsymbol{\beta}$.

The variance of the predicted utility is defined as
\begin{equation*}
    \mathrm{Var} \left[ \hat{U}_{js} \right] =
        \boldsymbol{f}^T(\boldsymbol{x}_{js}) \mathrm{Var} \left[ \boldsymbol{\hat{\beta}} \right] \boldsymbol{f}(\boldsymbol{x}_{js}) =
        \boldsymbol{f}^T(\boldsymbol{x}_{js}) \boldsymbol{I}^{-1}(\boldsymbol{X}, \boldsymbol{\beta}) \boldsymbol{f}(\boldsymbol{x}_{js}),
\end{equation*}
for a given parameter vector $\boldsymbol{\beta}$. As a result, the average variance of the predicted utility is
\begin{equation*}
    \int_{\chi} \boldsymbol{f}^T(\boldsymbol{x}_{js}) \boldsymbol{I}^{-1}(\boldsymbol{X}, \boldsymbol{\beta}) \boldsymbol{f}(\boldsymbol{x}_{js}) d\boldsymbol{x}_{js}
    =
    \int_{\chi} \mathrm{tr} \left[ \boldsymbol{I}^{-1}(\boldsymbol{X}, \boldsymbol{\beta}) \boldsymbol{f}(\boldsymbol{x}_{js}) \boldsymbol{f}^T(\boldsymbol{x}_{js}) \right] d\boldsymbol{x}_{js} 
    =
    \mathrm{tr} \left[ \boldsymbol{I}^{-1}(\boldsymbol{X}, \boldsymbol{\beta}) \boldsymbol{W}_u \right]
\end{equation*}
where 
\begin{equation}    \label{eq_moments_matrix_utility}
    \boldsymbol{W}_u = \int_{\chi}  \boldsymbol{f}(\boldsymbol{x}_{js}) \boldsymbol{f}^T(\boldsymbol{x}_{js}) d\boldsymbol{x}_{js}
\end{equation}
is the moments matrix corresponding to our definition of the I-optimality criterion, with the subscript $u$ referring to the fact that this is the moments matrix for the I-optimality criterion based on predicted utilities. 

In the event the experimental region $\chi$ is the $(q-1)$-dimensional simplex, there exists a closed-form expression for the moments matrix $\boldsymbol{W}_u$ that does not depend on the parameter vector $\boldsymbol{\beta}$ \cite{goos_i-optimal_2016}. This means that, even if a Bayesian optimal design is desired, the moments matrix needs to be computed only once in the design creation process, reducing the computational burden. The elements of the moments matrix $\boldsymbol{W}_u$ can be obtained using the following formula given in \citeauthor{goos_i-optimal_2016} \cite{goos_i-optimal_2016}, \citeauthor{ejor} \cite{ejor}, and \citeauthor{degroot2005optimal} \cite{degroot2005optimal}:
\begin{equation*}
    \int_{\chi} x_{1}^{p_1} x_{2}^{p_2} \hdots x_{q}^{p_q} dx_{1} dx_{2} \hdots dx_{q-1} 
        = \frac{\prod_{i = 1}^q \Gamma(p_i + 1)}{\Gamma(q + \sum_{i = 1}^q p_i)}.
\end{equation*}
Our local I-optimality criterion is given by
\begin{equation*}    \label{eq_i_optim_classical_2}
    \mathcal{I} = 
    \int_{\chi} \mathrm{Var} \left[ \hat{U}_{js} \right] d\boldsymbol{x}_{js}
     =
    \mathrm{tr}\left[  \boldsymbol{I}^{-1}(\boldsymbol{X}, \boldsymbol{\beta}) \boldsymbol{W}_u \right],
\end{equation*}
while our Bayesian I-optimality criterion is 
\begin{equation}    \label{eq_pseudo_bayes_i_eff_1}
    \mathcal{I}_B =
    \int_{\mathbb{R}^{r}}
    \mathrm{tr}\left[ \boldsymbol{I}^{-1}(\boldsymbol{X}, \boldsymbol{\beta}) \boldsymbol{W}_u\right] 
    \pi(\boldsymbol{\beta}) d\boldsymbol{\beta}.
\end{equation}
In these expressions, we ignore the volume of the experimental region, $1/(q-1)!$, because, for the optimization of the designs, it is an irrelevant constant.


\section{Construction of I-optimal designs}
\label{sec:construction_of_iopt_designs}

To compute Bayesian I-optimal designs for choice experiments with mixtures, we used a coordinate-exchange algorithm \cite{goos_jones_optimal_2011, meyer_nachtsheim}, modified to generate Bayesian I-optimal designs for choice experiments with mixtures. A coordinate-exchange algorithm was also used by \citeauthor{kessels_efficient_2009} \cite{kessels_efficient_2009} and \citeauthor{ruseckaite_bayesian_2017} \cite{ruseckaite_bayesian_2017} in the context of choice experimentation. Our algorithm was implemented in the R programming language \cite{rlang} in which we created a package called \texttt{opdesmixr} which is available at \texttt{https://github.com/mariobecerra/opdesmixr}. The package was created with the aid of several other R packages \cite{ggtern2018, ggplot2_2016, devtools2020, rcpp2011, rcpp2013, rcpp2018, rcppArmadillo2014, purrr2020}, and allows the computation of locally D-optimal, Bayesian D-optimal, locally I-optimal, and Bayesian I-optimal designs for first-order, second-order, and special-cubic Scheffé models. The user must specify either a single parameter vector for a locally optimal design or a matrix of draws from the parameter vector's prior distribution for a Bayesian optimal design.

Our coordinate-exchange algorithm starts from a random initial design, and starts by optimizing the first ingredient proportion of the first alternative within the first choice set, followed by the second ingredient proportion of the first alternative within the first choice set, and so on, until all $q$ ingredient proportions have been optimized. The algorithm then repeats this process for each alternative and each choice set in the design. The whole process is repeated until the design can no longer be improved or until a maximum number of iterations has been reached. As pointed out in \citeauthor{piepel_construction_2005} \cite{piepel_construction_2005}, \citeauthor{goos_jones_optimal_2011} \cite{goos_jones_optimal_2011}, and \citeauthor{ruseckaite_bayesian_2017} \cite{ruseckaite_bayesian_2017}, the coordinate-exchange algorithm has to undergo some modifications to deal with mixtures. As a matter of fact, because the mixture proportions have to sum up to one, they cannot be changed independently, and a change in one proportion requires a change in at least one other proportion. We deal with this dependency by changing proportions using the so-called Cox effect direction \cite{cornell2002experiments, goos_jones_optimal_2011, piepel_construction_2005}. This means that, after a change of one of the ingredient proportions, $x_{ijs}$, to $x_{ijs} + \Delta$, we modify the other $q-1$ proportions as follows:
\begin{equation*}
    x_{kjs}^{\mathrm{new}} := 
    \begin{cases} 
      \left( 1 - \frac{\Delta}{1 - x_{ijs}} \right) x_{kjs}  & \mathrm{if} \quad x_{ijs} \neq 1, \\
      \frac{ 1 - (x_{ijs} + \Delta)}{q - 1} & \mathrm{if} \quad x_{ijs} = 1. 
   \end{cases}
\end{equation*}

Three other aspects concerning our coordinate-exchange algorithm are worth mentioning too. First, we recommend running the coordinate-exchange algorithm multiple times, each time starting from a different random initial design. This is because the coordinate-exchange algorithm is a heuristic optimization algorithm, which cannot guarantee optimality, and by running it multiple times, we have a bigger chance of finding a truly optimal design. The larger the number of ingredients and the more complex the model, the larger the number of starts of the coordinate-exchange algorithm should be. For the examples discussed in Section \ref{sec:results}, we use 80 random starts of the algorithm because the number of ingredients is as low as three and the number of parameters is only six. Second, we seek the optimal value of every individual ingredient proportion $x_{ijs}$ using Brent's univariate optimization method \cite{brent1973algorithms}. Third, we need to approximate the Bayesian optimality criteria numerically, because there is no closed-form solution to the integrals in Equations~(\ref{eq_pseudo_bayes_d_eff}) and~(\ref{eq_pseudo_bayes_i_eff_1}). It is common to do this utilizing random or systematic draws from the prior distribution $\pi(\boldsymbol{\beta})$ \cite{train_discrete_2009, ruseckaite_bayesian_2017, kessels_efficient_2009, yu2010comparing}. Denoting the $R$ draws from the prior distribution by $\boldsymbol{\beta}^{(i)}$, the approximations for Equations (\ref{eq_pseudo_bayes_d_eff}) and (\ref{eq_pseudo_bayes_i_eff_1}) are 
\begin{equation}    \label{eq_pseudo_bayes_d_eff_mc}
    \mathcal{D}_B \approx 
        \log{\left( 
        \frac{1}{R} \sum_{i = 1}^R
        \left[  
        \det \left( 
            \boldsymbol{I}^{-1}(\boldsymbol{X}, \boldsymbol{\beta}^{(i)})
         \right) \right]^{\frac{1}{r}} 
        \right)},
\end{equation}
and
\begin{equation}    \label{eq_pseudo_bayes_i_eff_mc}
    \mathcal{I}_B \approx
        \frac{1}{R} \sum_{i = 1}^R
        \mathrm{tr} \left[ \boldsymbol{I}^{-1}(\boldsymbol{X}, \boldsymbol{\beta}^{(i)}) \boldsymbol{W}_u \right].
\end{equation}
One commonly used method to obtain draws from a prior distribution involves Halton sequences. The resulting Halton draws provide a good coverage of
the entire density domain, as well as negatively correlated draws that reduce the variance of the approximation to the integral \cite{train_discrete_2009}. Therefore, like \citeauthor{ruseckaite_bayesian_2017} \cite{ruseckaite_bayesian_2017}, we used 128 Halton draws from the prior distribution in both of our examples in the next section. The number 128 provides a good enough approximation for the number of parameters in the models used in the two examples. For choice experiments involving more model parameters, a larger number of Halton draws should be used. For more details about Halton draws and other approximation methods, as well as their advantages and disadvantages, we refer to \citeauthor{yu2010comparing} \cite{yu2010comparing}.


\section{Results}\label{sec:results}

This section shows our computational results for two example choice experiments involving a mixture. The first example is the taste experiment involving cocktails from \citeauthor{courcoux1997methode} \cite{courcoux1997methode}, while the second example involves an experiment with artificial sweeteners for a sports drink from \citeauthor{cornell2002experiments} \cite{cornell2002experiments}.

\subsection{Cocktail preferences}

\citeauthor{ruseckaite_bayesian_2017} \cite{ruseckaite_bayesian_2017} revisited an experiment by \citeauthor{courcoux1997methode} \cite{courcoux1997methode} in which seven fruit cocktails involving mango juice (whose true proportion we denote by $a_1$), blackcurrant syrup (whose true proportion we denote by $a_2$), and lemon juice (whose true proportion we denote by $a_3$) were tasted. This was done by 60 consumers which were asked to taste different pairs of the seven fruit cocktails and to indicate their preferred cocktail in each pair. Each respondent had to evaluate eight of 21 possible pairs, resulting in a final experimental design with $60 \times 8 = 480$ choice sets of size 2.

In the experiment, \citeauthor{courcoux1997methode} \cite{courcoux1997methode} imposed lower bounds of 0.3, 0.15 and 0.1 on the three true ingredient proportions $a_1$, $a_2$, and $a_3$. To deal with this issue and to be able to use our implementation of the coordinate-exchange algorithm, like \citeauthor{ruseckaite_bayesian_2017} \cite{ruseckaite_bayesian_2017} did, we expressed the mixtures defining the cocktails in terms of so-called pseudocomponents $x_1$, $x_2$, and $x_3$. These pseudocomponents are defined such that they take a minimum value of $0$ and a maximum value of $1$, and sum up to one. The conversion of the true ingredient proportions into pseudocomponent proportions is done via the formula $x_i = (a_i - L_i)/(1 - L)$, where $L_i$ denotes the lower bound of ingredient $i$ and $L$ is the sum of the lower bounds for all $q$ ingredient proportions. 

To compute D-optimal designs for the cocktail experiment, \citeauthor{ruseckaite_bayesian_2017} \cite{ruseckaite_bayesian_2017} obtained a prior distribution for the parameter vector $\boldsymbol{\beta}$ in a special-cubic Scheffé model. More specifically, they re-analyzed the data from \citeauthor{courcoux1997methode} \cite{courcoux1997methode} and derived a multivariate normal prior distribution for $\boldsymbol{\beta}$ with mean vector $\boldsymbol{\beta}_0 = (1.36, 1.57, 2.47, -0.43, 0.50, 1.09)^T$ and variance-covariance matrix
\begin{equation*}
    \boldsymbol{\Sigma}_0 = 
    \begin{pmatrix*}[r]
    6.14  & 5.00  & 2.74  & -0.43 & -2.81 & -3.33 \\
    5.00  & 6.76  & 4.47  & -1.79 & -6.13 & -3.51 \\
    2.74  & 4.47  & 3.45  & -1.38 & -4.71 & -2.17 \\
    -0.43 & -1.79 & -1.38 & 1.18  & 2.39  & 0.71 \\
    -2.81 & -6.13 & -4.71 & 2.39  & 7.43  & 2.71 \\
    -3.33 & -3.51 & -2.17 & 0.71  & 2.71  & 2.49
    \end{pmatrix*}.
\end{equation*}

We used the same prior distribution to compute optimal designs with 16 choice sets of size 2 assuming a special-cubic Scheffé model. First, we computed a Bayesian D-optimal design to benchmark our implementation of the coordinate-exchange algorithm against that of \citeauthor{ruseckaite_bayesian_2017} \cite{ruseckaite_bayesian_2017}, and observed that our design has a slightly better D-optimality criterion value than the original when evaluated using our set of Halton draws. After this validation of our algorithm, we also computed a Bayesian I-optimal design. Our Bayesian D- and I-optimal designs are given in Tables \ref{tab:cocktail_exp_d_optimal_des} and \ref{tab:cocktail_exp_i_optimal_des} in the appendix and shown graphically in Figure \ref{fig:res_cocktail_db_vs_ib_design_simplex}. In the figure, the mixtures in each of the 16 choice sets are presented in terms of the pseudocomponent proportions and visualized using different markers for each choice set. The four colored areas in the graph correspond to four prior utility intervals, corresponding to the cutoff values $0$, $0.5625$, $1.125$, $1.6875$, and $2.25$. The yellow area is the set of mixtures with the lowest a priori utility values, while the red area indicates the set of mixtures with the highest a priori utility values. 

\begin{figure}[ht]
  \centering
    \begin{subfigure}[b]{0.45\textwidth}
      \centering
      \includegraphics[width=\textwidth]{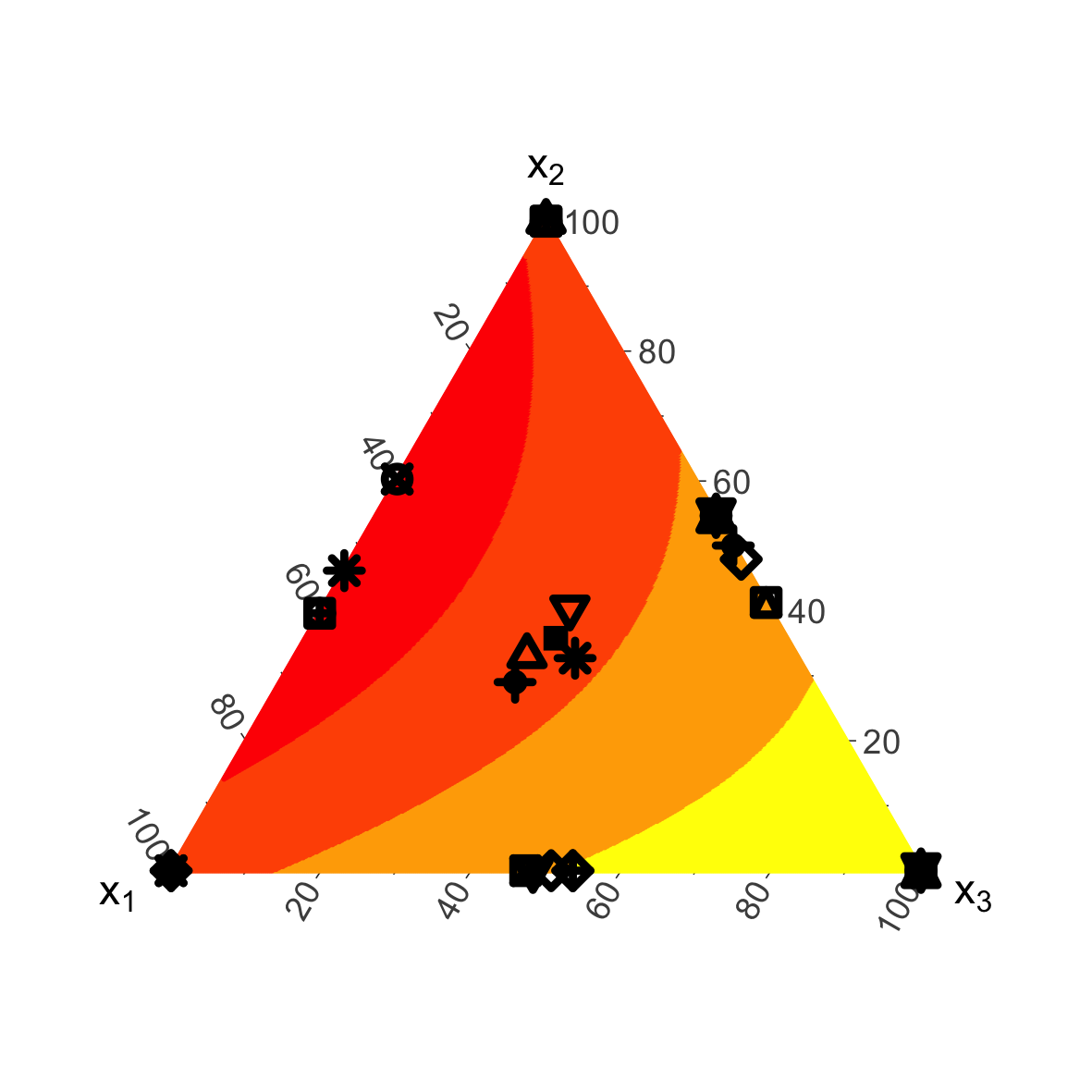}
      \caption{Bayesian D-optimal design}
      \label{fig:cocktail_des_subfig_db}
    \end{subfigure}
    \begin{subfigure}[b]{0.45\textwidth}
      \centering
      \includegraphics[width=\textwidth]{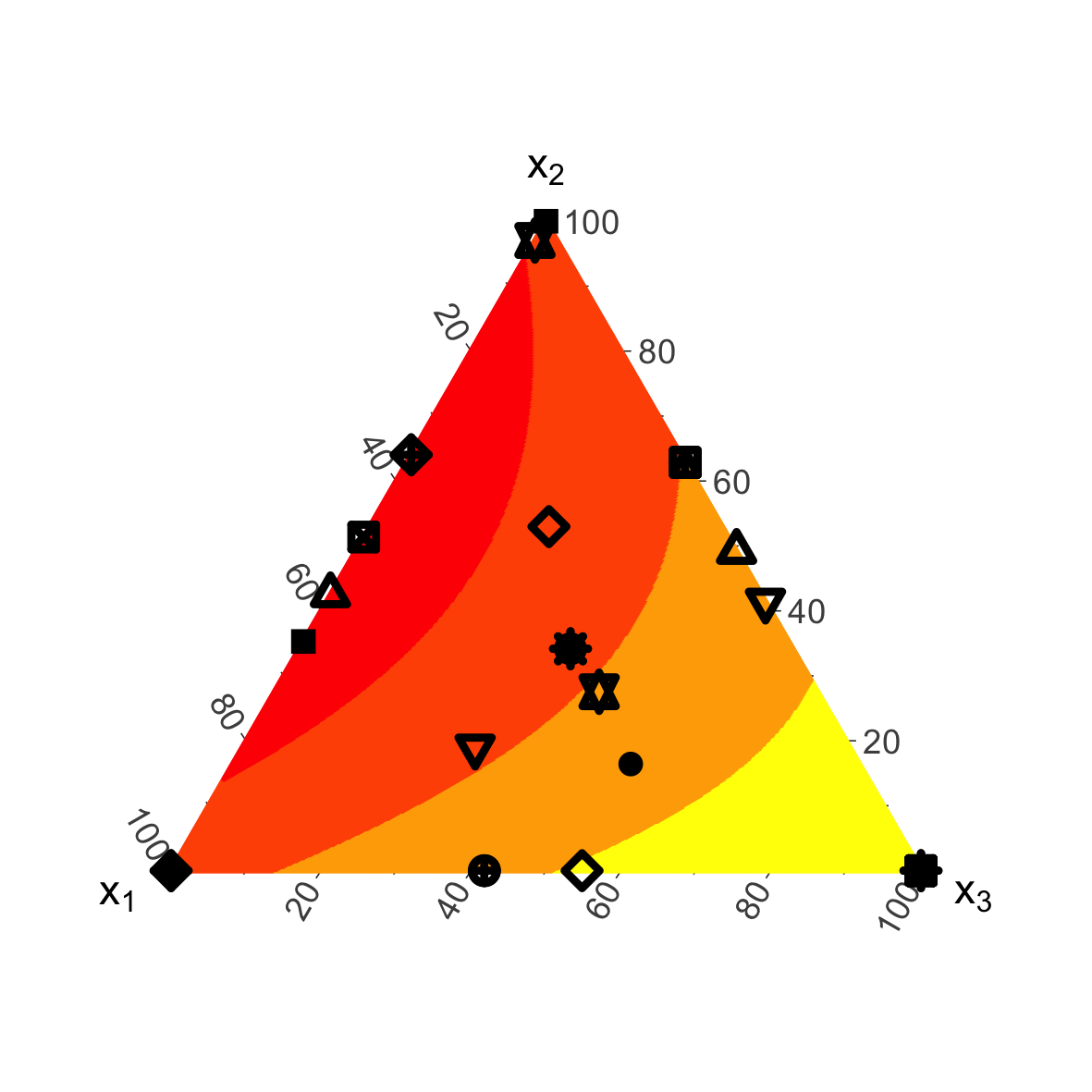}
      \caption{Bayesian I-optimal design}
      \label{fig:cocktail_des_subfig_ib}
    \end{subfigure}
  \caption{Bayesian optimal designs produced by our coordinate-exchange algorithm for the cocktail experiment. The colors represent utilities belonging to the following intervals:
  \legendsquare{yellow_mb}~$[0,0.5625)$,
    \legendsquare{yellow_red_mb_1}~$[0.5625,1.125)$,
    \legendsquare{yellow_red_mb_2}~$[1.125,1.6875)$,
    \legendsquare{red_mb}~$[1.6875,2.25)$.
}
  \label{fig:res_cocktail_db_vs_ib_design_simplex}
\end{figure}

\begin{figure}[ht]
    \centering
    \includegraphics[width=0.8\textwidth]{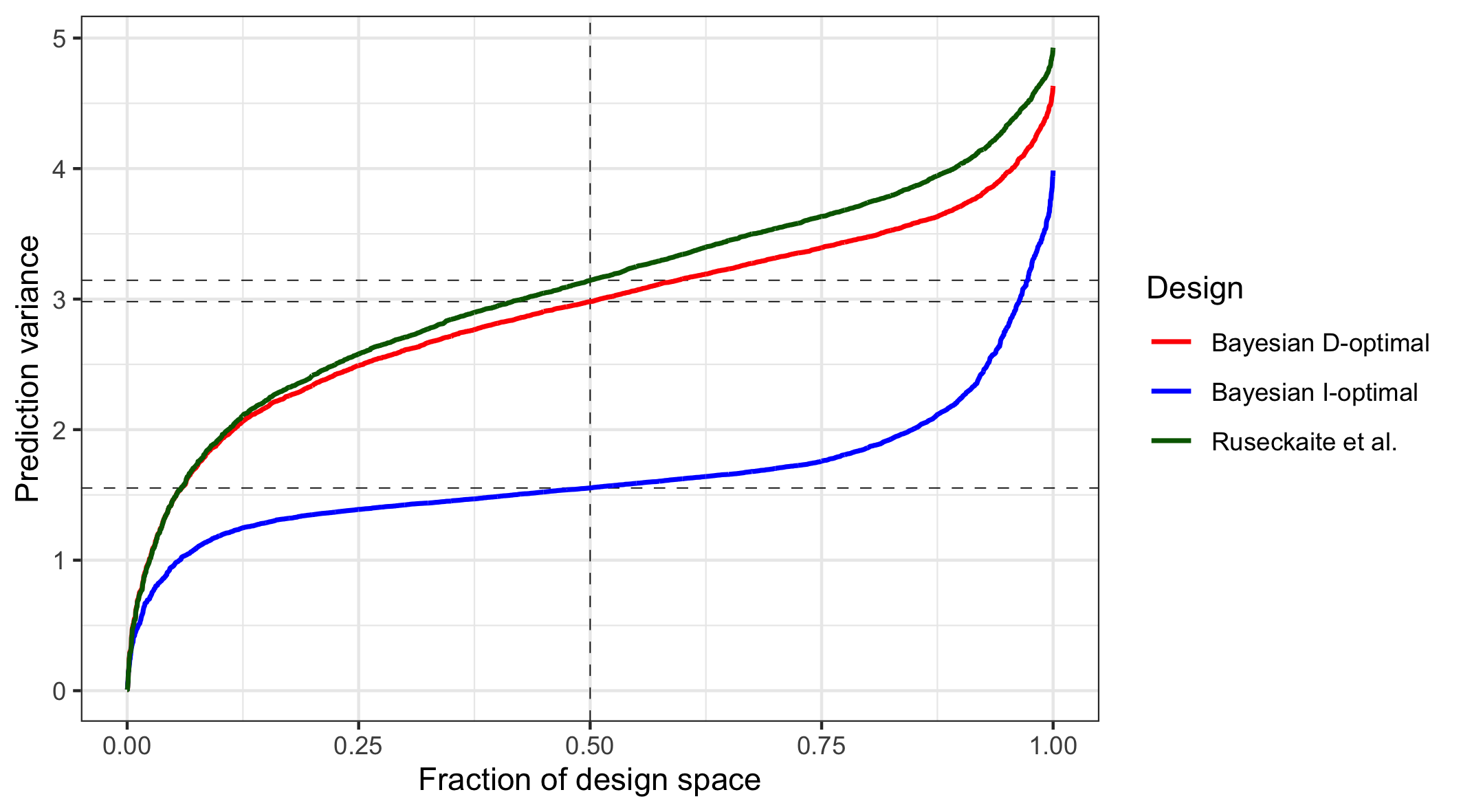}
    \caption{Fraction of design space plot of our Bayesian D- and I-optimal designs for the cocktail experiment as well as the Bayesian D-optimal design from \citeauthor{ruseckaite_bayesian_2017} \cite{ruseckaite_bayesian_2017}.}
    \label{fig:res_cocktail_fds_db_vs_ib_plot}
\end{figure}

Figure \ref{fig:res_cocktail_fds_db_vs_ib_plot} shows the fraction of the design space plots of the two Bayesian optimal designs we computed and of the benchmark design from \citeauthor{ruseckaite_bayesian_2017} \cite{ruseckaite_bayesian_2017}. Fraction of design space plots were originally introduced by \citeauthor{zahran2003fraction} \cite{zahran2003fraction} and display the performance of a design in terms of the prediction variance for each point in the experimental region or design space. The horizontal axis corresponds to a fraction of the experimental region, while the vertical axis ranges from the minimum prediction variance to the maximum prediction variance over the entire experimental region \cite{goos_jones_optimal_2011}. In the context of choice experiments, the prediction variance depends on the unknown parameter vector. We dealt with this issue by computing prediction variances for $128$ Halton draws from the prior distribution of the parameter vector and averaging the results.

The most striking conclusion from the fraction of design space plot is that the prediction variances are much lower for the Bayesian I-optimal design than for the Bayesian D-optimal designs. The plot shows, for instance, that the median prediction variance for the Bayesian I-optimal design is about $1.55$, while it is about $3$ for our Bayesian D-optimal design and $3.2$ for the benchmark design. The maximum prediction variance is also substantially lower for the Bayesian I-optimal design. In summary, using the Bayesian I-optimal design provides much added value in terms of precision of prediction when compared to Bayesian D-optimal designs.

\begin{figure}[ht]
  \centering
    \begin{subfigure}[b]{0.35\textwidth}
      \centering
      \includegraphics[width=\textwidth]{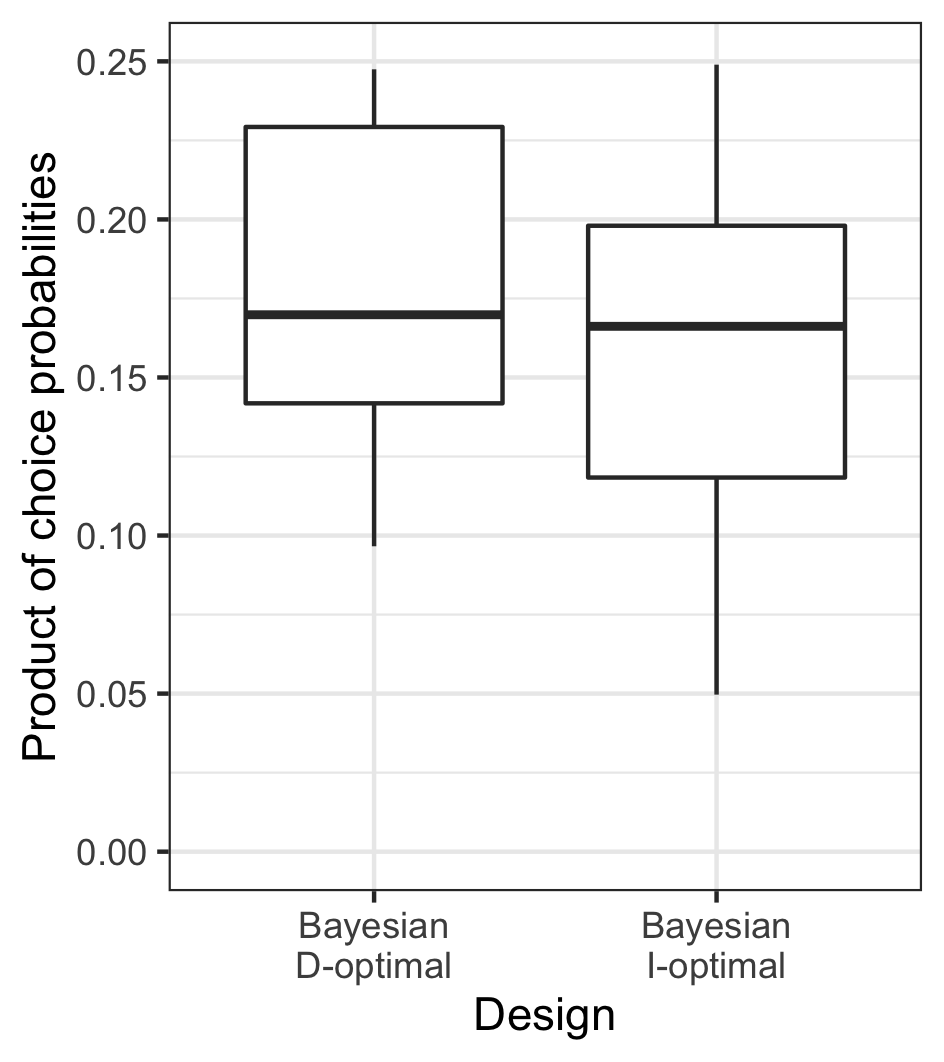}
      \caption{Utility balance}
      \label{fig:res_cocktail_choice_probs_plot}
    \end{subfigure}
    \hspace{1cm}
    \begin{subfigure}[b]{0.35\textwidth}
      \centering
      \includegraphics[width=\textwidth]{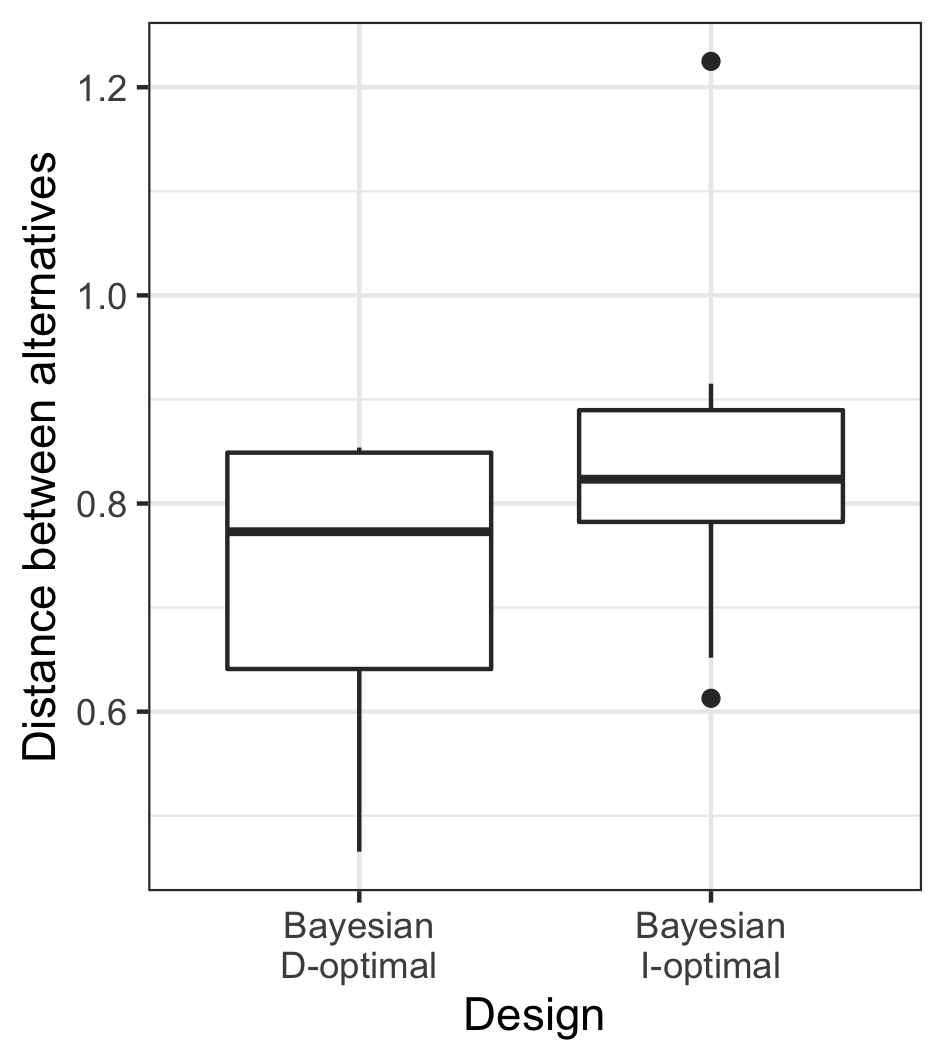}
      \caption{Euclidean distances}
      \label{fig:res_cocktail_distances_within_choice_set}
    \end{subfigure}
  \caption{Technical properties of our Bayesian D- and I-optimal designs for the cocktail experiment}
  \label{fig:res_cocktail_choice_probs_dist}
\end{figure}

It is not easy to describe the properties of optimal choice designs. To compare the Bayesian D- and I-optimal designs, we quantified the utility balance in the designs' choice sets and computed the Euclidean distances between the alternatives in the choice sets. Utility balance refers to the property that alternatives within a choice set possess the same or a similar a priori utility and therefore have the same or almost the same a priori choice probability. Utility balance was advocated by \citeauthor{huber1996importance} \cite{huber1996importance} as a desirable property for choice designs. A choice set of two alternatives is perfectly utility balanced in the event the choice probabilities of the two alternatives both equal $0.5$ and the product of the two probabilities is $0.25$. Choice sets that are not at all utility balanced involve alternatives with very different utilities and choice probabilities. For such choice sets, the product of the two choice probabilities is substantially lower than $0.25$.
Figure \ref{fig:res_cocktail_choice_probs_plot} shows boxplots of the product of the choice probabilities in the choice sets of our Bayesian D-optimal and our Bayesian I-optimal design. The D-optimal design tends to have choice sets with a higher utility balance than the I-optimal design. However, the median values for the products of the choice probabilities in both designs are very similar and roughly equal to 0.17. This value corresponds to choice probabilities of about 0.78 and 0.22, implying that the Bayesian optimal designs do not exhibit much utility balance.
Figure \ref{fig:res_cocktail_distances_within_choice_set} shows the Euclidean distances between the two alternatives within a choice set for our Bayesian D- and I-optimal designs. The alternatives within a single choice set tend to be closer together in the D-optimal design than in the I-optimal design. This can also be observed in Figure \ref{fig:res_cocktail_db_vs_ib_design_simplex} by comparing the distances between the mixtures represented by a given marker. That the alternatives within the choice sets of the I-optimal design are located further from each other is in line with our observations that they exhibit less utility balance. As a matter of fact, when the distance between two mixtures is large, they typically appear in a different utility interval (in other words, in a differently colored area) and their choice probabilities are necessarily quite different.

\subsection{Artificial sweetener experiment}

As a second example, we also revisit the artificial sweetener experiment, a three-ingredient mixture experiment from \citeauthor{cornell2002experiments} \cite{cornell2002experiments}, intended to investigate whether an artificial sweetener could be used in an athletic sports drink. The original response variable of interest was ‘intensity of sweetness aftertaste’, but, just like \citeauthor{ruseckaite_bayesian_2017} \cite{ruseckaite_bayesian_2017}, we interpret the intensity of sweetness aftertaste as being proportional to the utility in the multinomial logit model and consider a choice experiment consisting of 7 choice sets of size two as an alternative to the original experiment in \cite{cornell2002experiments}.

For this example, \citeauthor{ruseckaite_bayesian_2017} \cite{ruseckaite_bayesian_2017} started with a special-cubic Scheffé model to construct Bayesian D-optimal designs with a multivariate normal prior distribution with mean vector
\begin{equation*}
    \boldsymbol{\beta}_0 = (0.86, 0.21, 3.07, 2.34, 3.24, -20.59)^T
\end{equation*}
and variance-covariance matrices of the form $\boldsymbol{\Sigma}_0 = \kappa \boldsymbol{I}_7$, where $\kappa$ is a positive scalar that controls the level of uncertainty and $\boldsymbol{I}_7$ is the identity matrix of size $7$. A higher value of $\kappa$ indicates a higher level of uncertainty concerning the parameter values. The variance-covariance matrix was transformed to the identified parameter space, as mentioned in Section~\ref{subsec:mixtures_and_mnl_model}. The transformed variance-covariance matrix, denoted by $\boldsymbol{\Sigma}_0^{'}$, is
\begin{equation*}
    \boldsymbol{\Sigma}_0^{'} = 
    \begin{pmatrix*}[r]
    2 \kappa    & \kappa  & 0       & 0      & 0          & 0 \\
    \kappa      & 2\kappa & 0       & 0      & 0          & 0 \\
    0           & 0       & \kappa  & 0      & 0          & 0 \\
    0           & 0       & 0       & \kappa & 0          & 0 \\
    0           & 0       & 0       & 0      & \kappa     & 0 \\
    0           & 0       & 0       & 0      & 0          & \kappa
    \end{pmatrix*}.
\end{equation*}

\begin{figure}[ht]
     \captionsetup[subfigure]{aboveskip=-3pt,belowskip=-2pt}
    \centering
    
  \begin{subfigure}[b]{0.33335\textwidth}
    \centering
    \includegraphics[width=\textwidth]{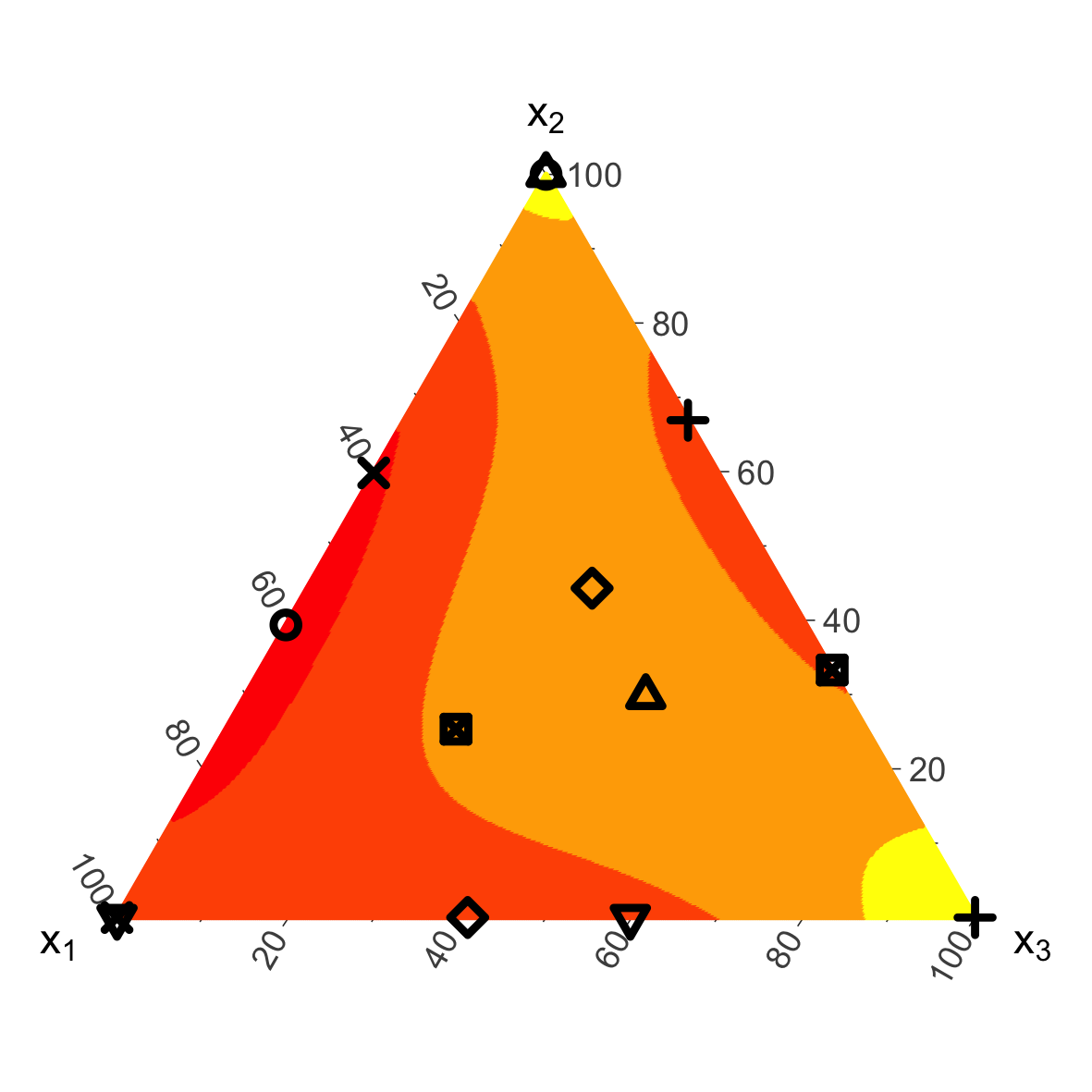}
    \caption{D-optimal design with $\kappa = 0.5$}
    \label{fig:cornell_simplex_subfig_kappa_005_D_opt}
  \end{subfigure}
  \begin{subfigure}[b]{0.33335\textwidth}
    \centering
    \includegraphics[width=\textwidth]{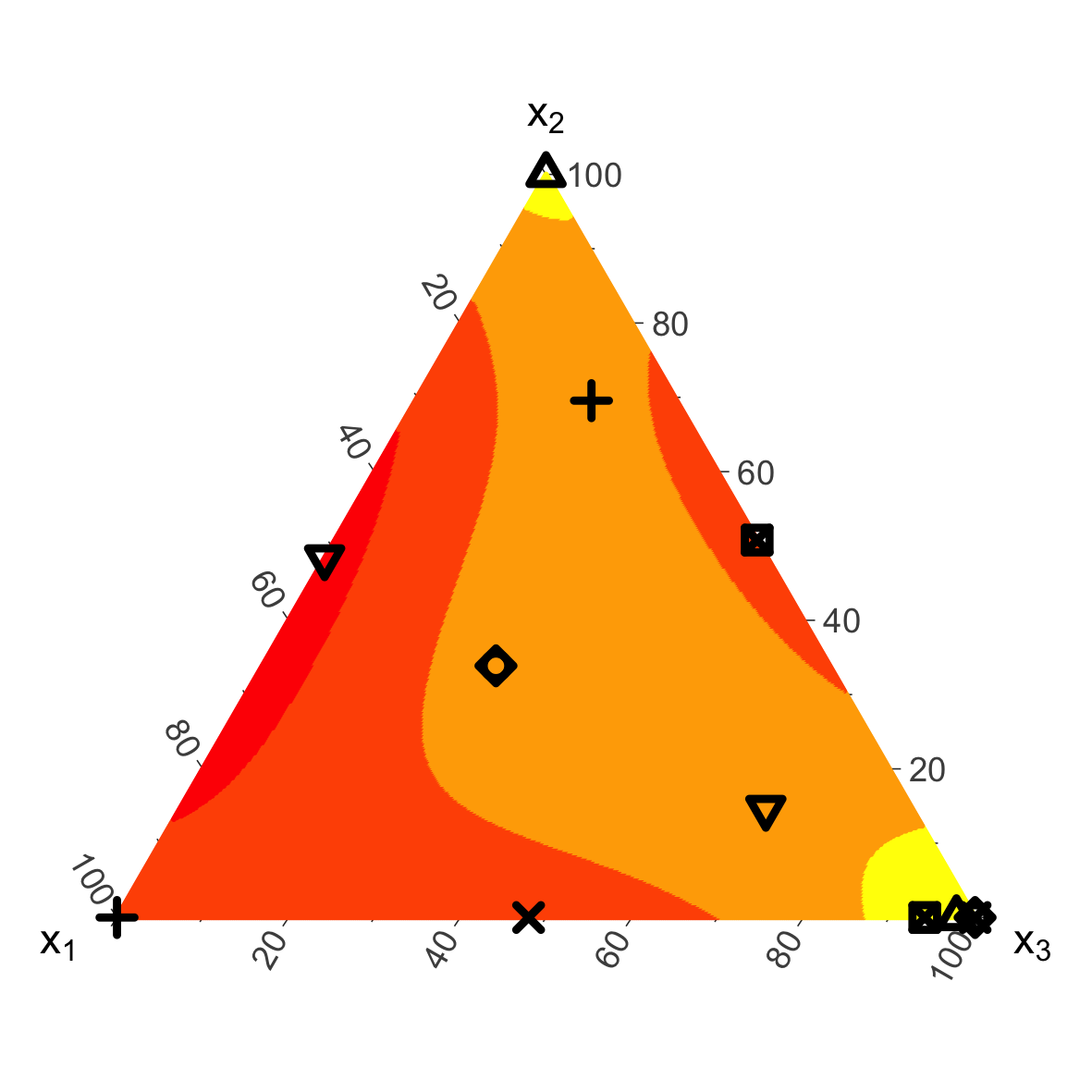}
    \caption{I-optimal design with $\kappa = 0.5$}
    \label{fig:cornell_simplex_subfig_kappa_005_I_opt}
  \end{subfigure}
  \hfill
  \begin{subfigure}[b]{0.33335\textwidth}
    \centering
    \includegraphics[width=\textwidth]{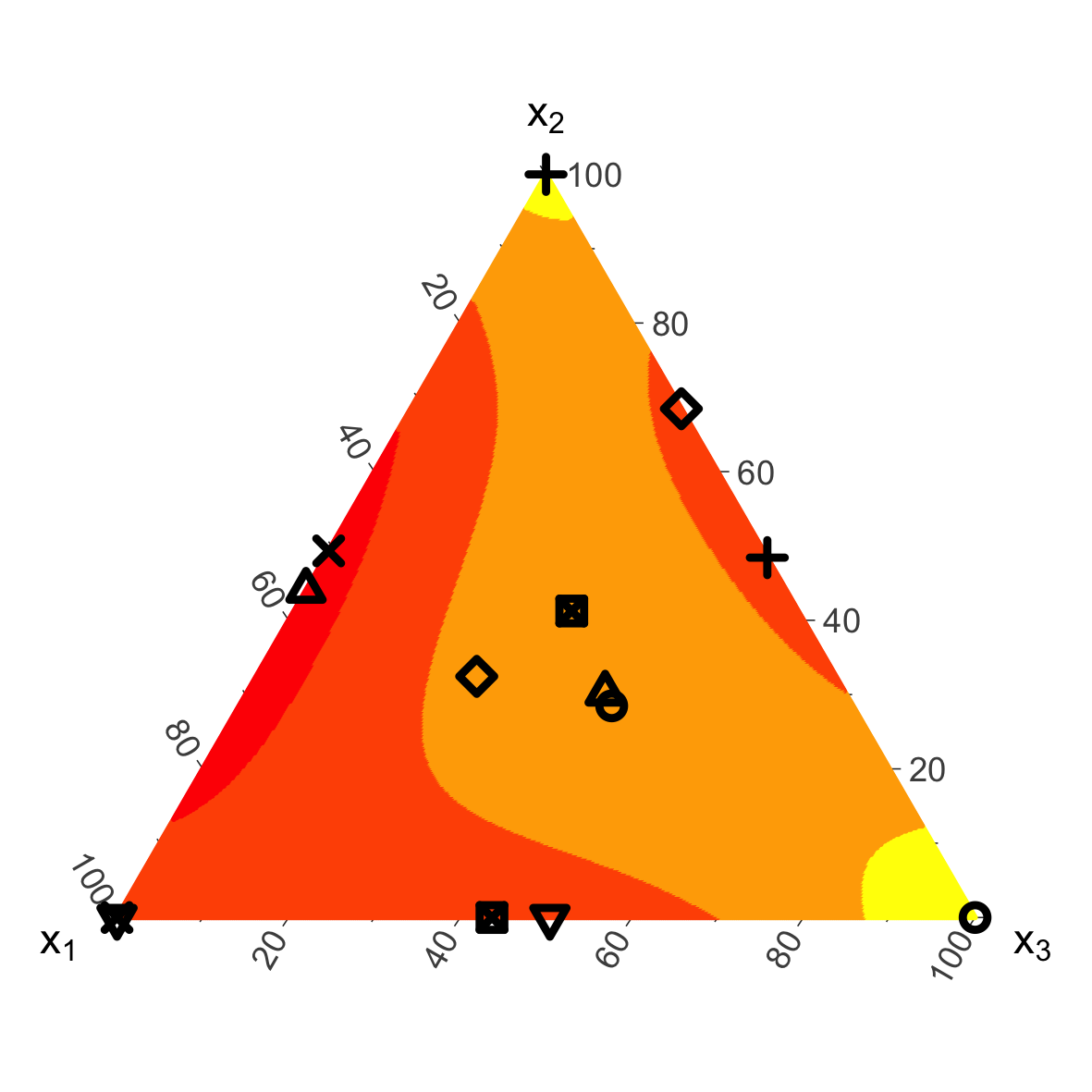}
    \caption{D-optimal design with $\kappa = 5$}
    \label{fig:cornell_simplex_subfig_kappa_050_D_opt}
  \end{subfigure}
  \begin{subfigure}[b]{0.33335\textwidth}
    \centering
    \includegraphics[width=\textwidth]{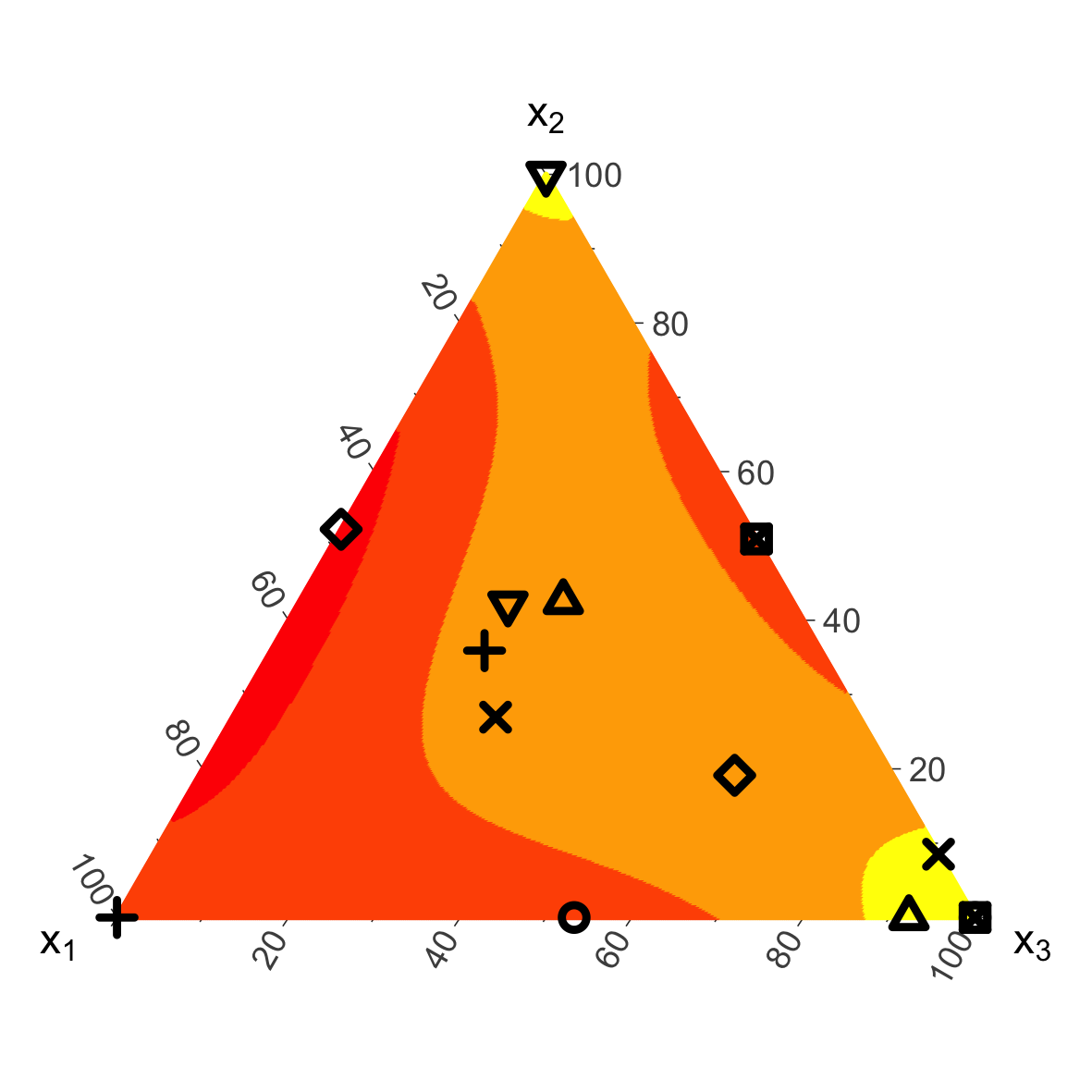}
    \caption{I-optimal design with $\kappa = 5$}
    \label{fig:cornell_simplex_subfig_kappa_050_I_opt}
  \end{subfigure}
  \hfill
  \begin{subfigure}[b]{0.33335\textwidth}
    \centering
    \includegraphics[width=\textwidth]{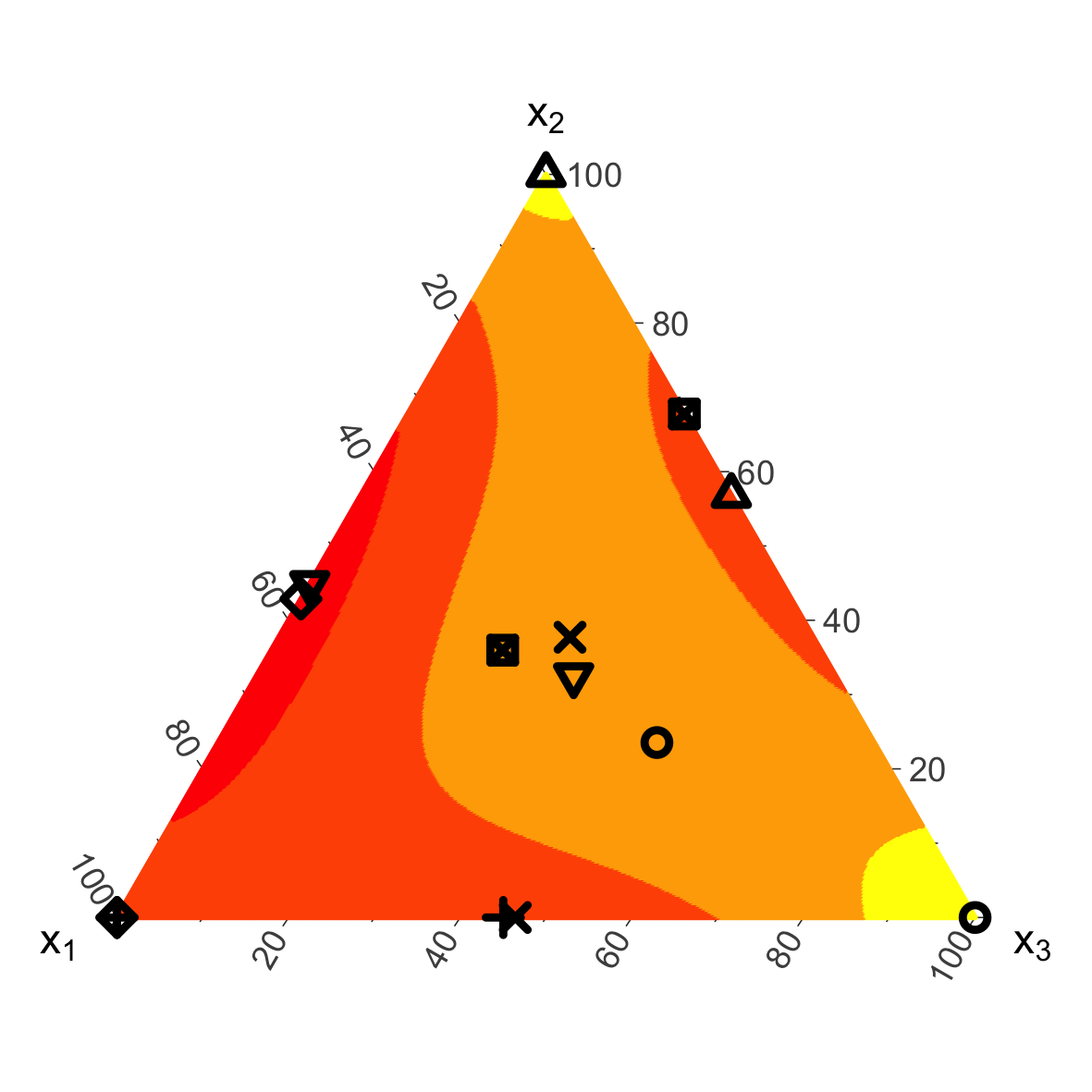}
    \caption{D-optimal design with $\kappa = 10$}
    \label{fig:cornell_simplex_subfig_kappa_100_D_opt}
  \end{subfigure}
  \begin{subfigure}[b]{0.33335\textwidth}
    \centering
    \includegraphics[width=\textwidth]{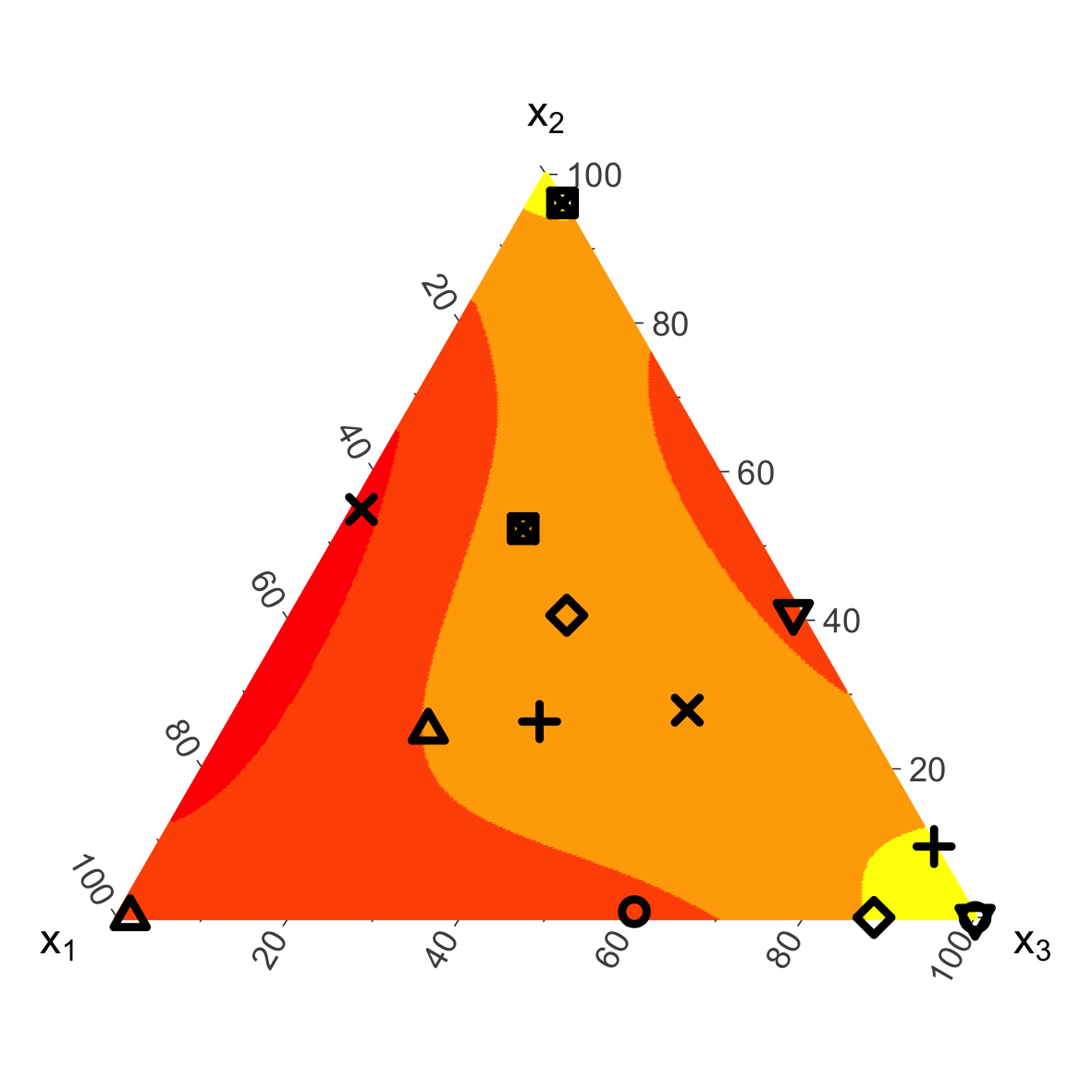}
    \caption{I-optimal design with $\kappa = 10$}
    \label{fig:cornell_simplex_subfig_kappa_100_I_opt}
  \end{subfigure}
  \hfill
  \begin{subfigure}[b]{0.33335\textwidth}
    \centering
    \includegraphics[width=\textwidth]{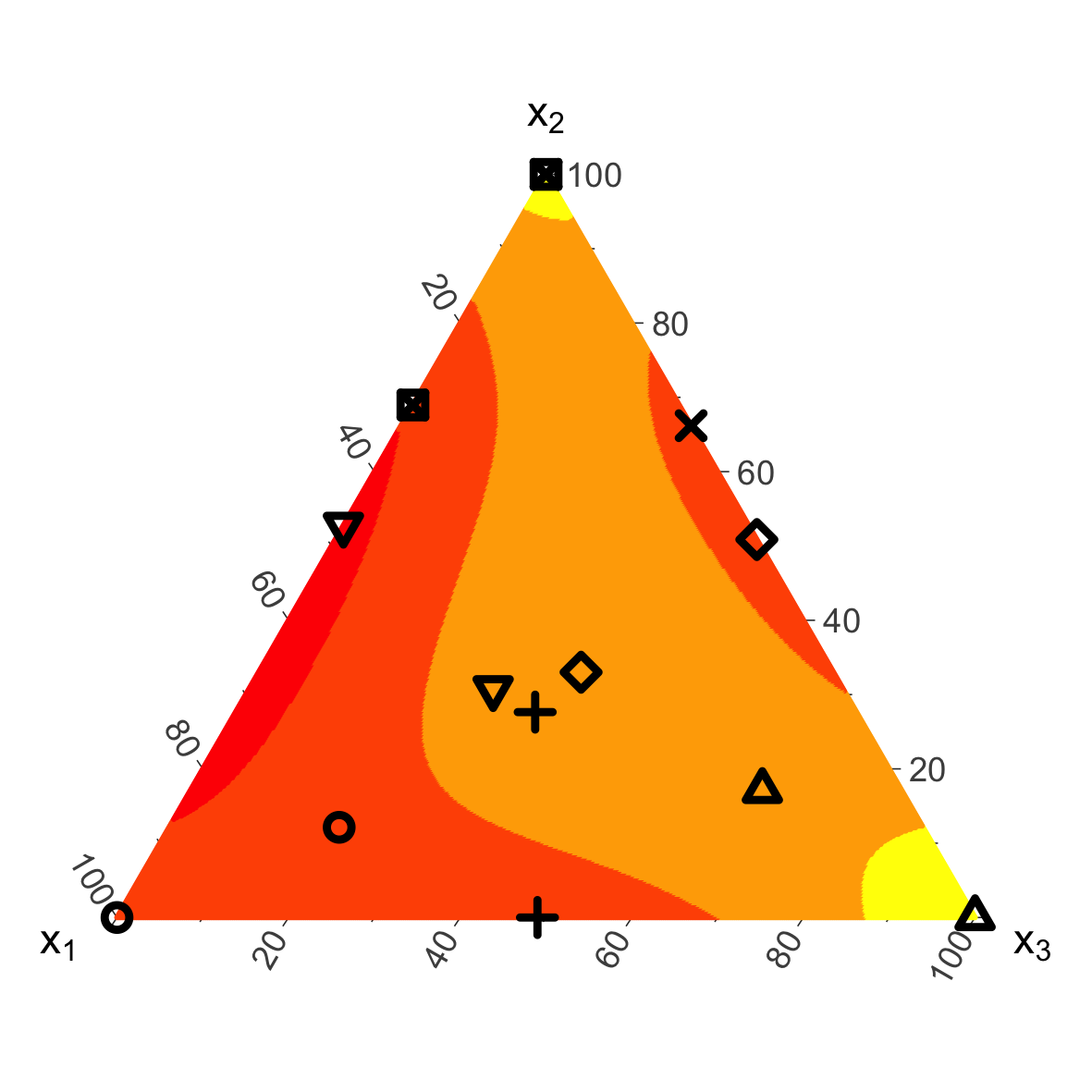}
    \caption{D-optimal design with $\kappa = 30$}
    \label{fig:cornell_simplex_subfig_kappa_300_D_opt}
  \end{subfigure}
  \begin{subfigure}[b]{0.33335\textwidth}
    \centering
    \includegraphics[width=\textwidth]{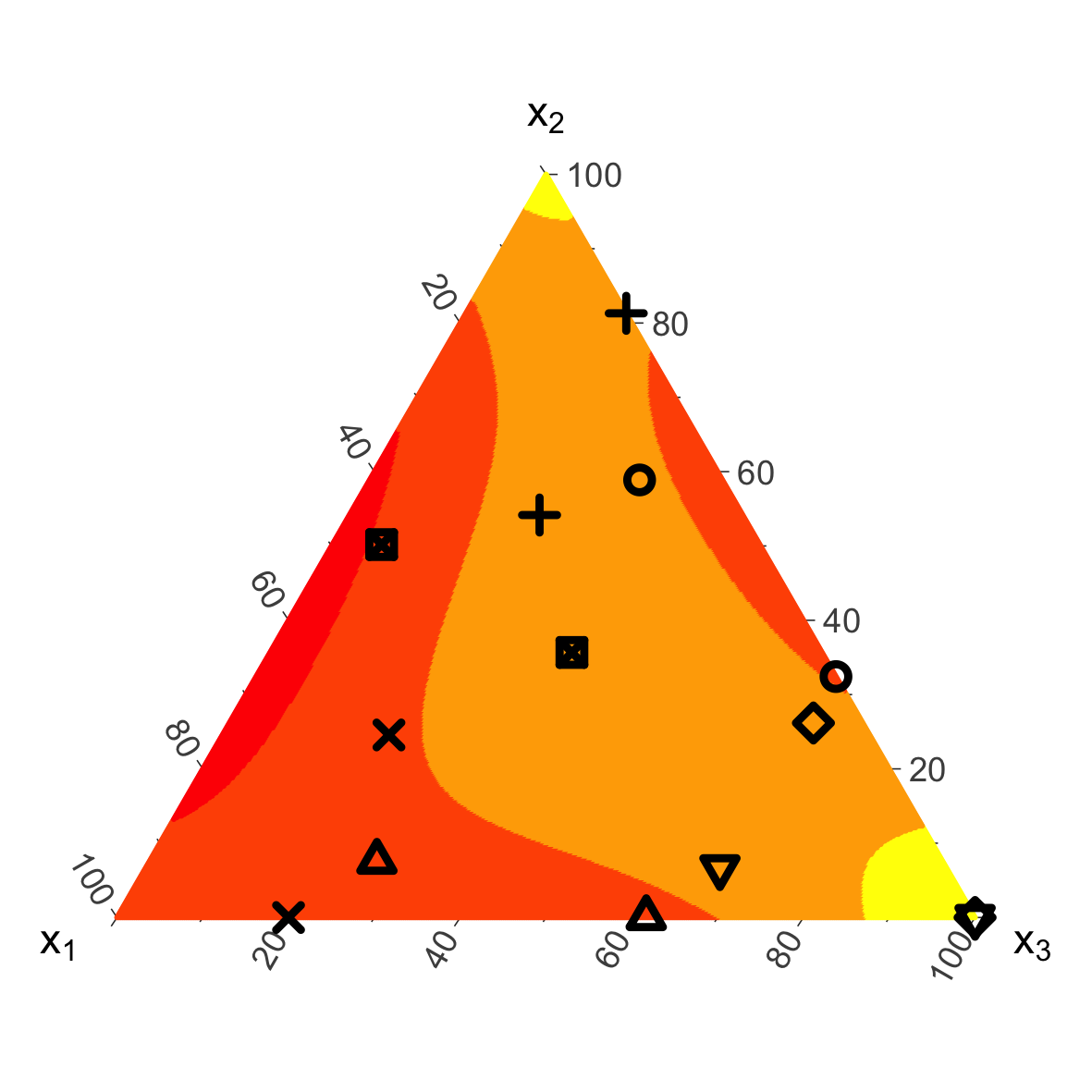}
    \caption{I-optimal design with $\kappa = 30$}
    \label{fig:cornell_simplex_subfig_kappa_300_I_opt}
  \end{subfigure}
  \hfill
  \caption{
  Bayesian optimal designs for the artificial sweetener experiment. The colors represent utilities belonging to the following intervals:
  \legendsquare{yellow_mb}~$[0,0.375)$,
    \legendsquare{yellow_red_mb_1}~$[0.375,0.75)$,
    \legendsquare{yellow_red_mb_2}~$[0.75,1.125)$,
    \legendsquare{red_mb}~$[1.125,1.34)$.
    }
  \label{fig:res_cornell_simplex}
\end{figure}

With our implementation of the coordinate-exchange algorithm, we first computed Bayesian D-optimal designs for the same $\kappa$ values as \citeauthor{ruseckaite_bayesian_2017} \cite{ruseckaite_bayesian_2017}, namely $0.5$, $5$, $10$ and $30$. When comparing our D-optimal designs to those of \citeauthor{ruseckaite_bayesian_2017} \cite{ruseckaite_bayesian_2017} using our sets of 128 Halton draws from the prior distributions, we observed that our designs have D-optimality criterion values very close to those of \citeauthor{ruseckaite_bayesian_2017} \cite{ruseckaite_bayesian_2017}, with three designs being slightly better and one being slightly worse. This provides another validation to our algorithm. We also computed Bayesian I-optimal designs for the four $\kappa$ values. All of our Bayesian D- and I-optimal designs are shown graphically in Figure~\ref{fig:res_cornell_simplex}. They are given in tabular format in Tables~\ref{tab:cornell_exp_d_optimal_des_kappa_0.5}--\ref{tab:cornell_exp_i_optimal_des_kappa_30} in the appendix. The four different colors in Figure \ref{fig:res_cornell_simplex} correspond to four intervals for the utility of the mixtures, with bounds $0$, $0.375$, $0.75$, $1.125$, and $1.34$. It can be seen that the spread in the points in the optimal designs increases with $\kappa$, but this phenomenon is more pronounced for the Bayesian I-optimal designs than for the Bayesian D-optimal designs.

\begin{figure}[ht]
    \centering
    \includegraphics[width=0.85\textwidth]{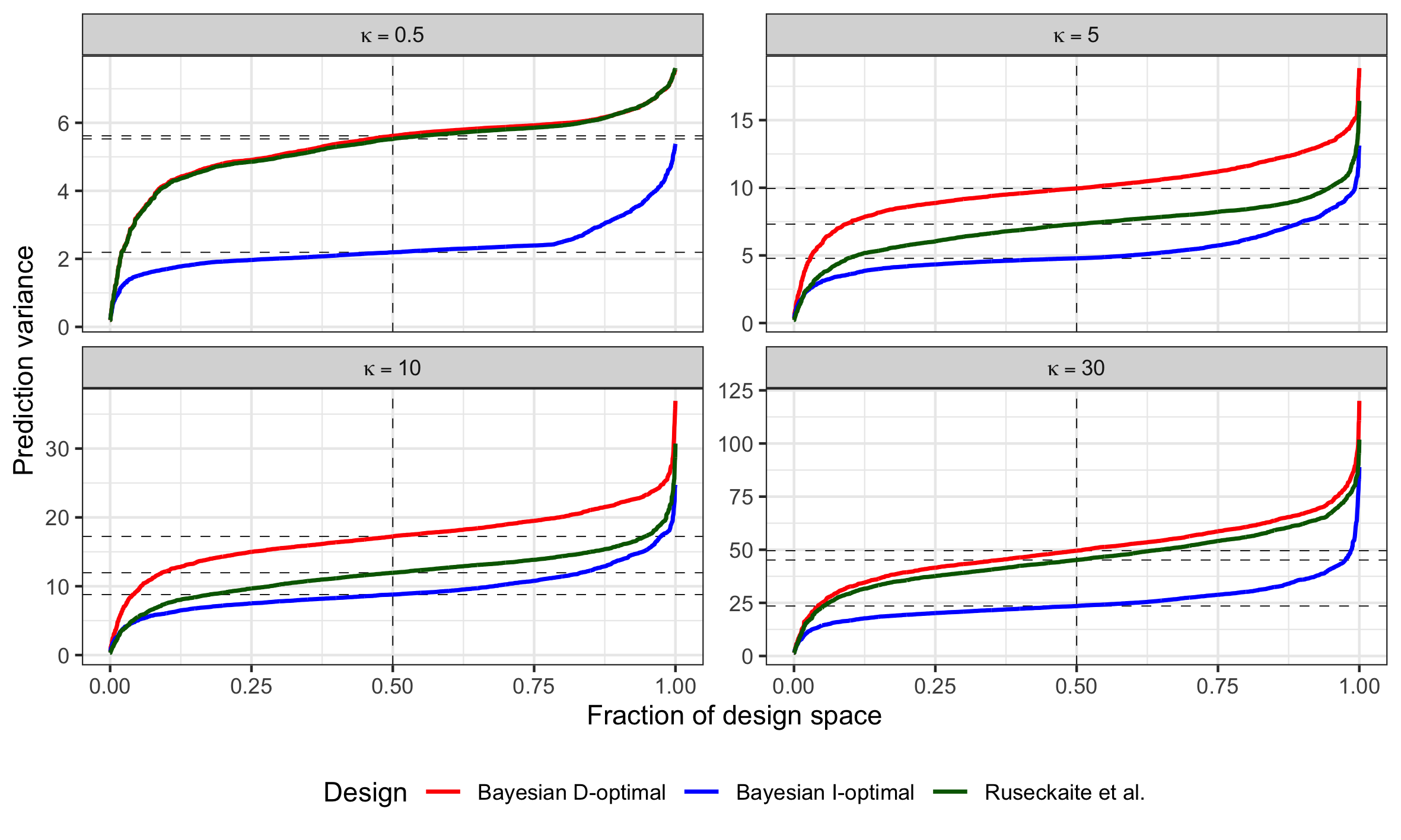}
    \caption{Fraction of design space plots of the Bayesian optimal designs for the artificial sweetener experiment}
    \label{fig:res_cornell_db_vs_ib_fds_plot}
\end{figure}

Figure \ref{fig:res_cornell_db_vs_ib_fds_plot} shows the fraction of design space plots for our Bayesian D- and I-optimal designs, as well as for the designs of \citeauthor{ruseckaite_bayesian_2017} \cite{ruseckaite_bayesian_2017}. For each value of $\kappa$, the I-optimal design has a much lower prediction variance than our D-optimal design and that of \citeauthor{ruseckaite_bayesian_2017} \cite{ruseckaite_bayesian_2017}. The difference in predictive performance, in favor of the Bayesian I-optimal designs, increases with $\kappa$.

\begin{figure}[ht]
    \centering
    \includegraphics[width=0.9\textwidth]{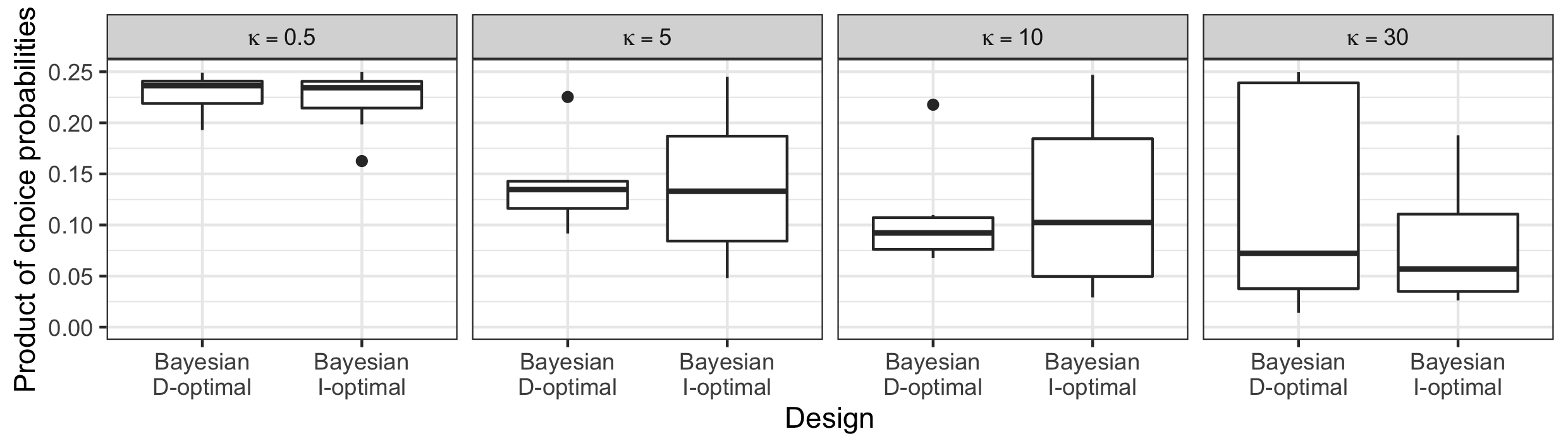}
    \caption{Measures of utility balance for our Bayesian D- and I-optimal designs for the artificial sweetener experiment}
    \label{fig:res_cornell_analytic_transf_choice_probs_plot}
\end{figure}

Figure \ref{fig:res_cornell_analytic_transf_choice_probs_plot} shows boxplots of the product of the choice probabilities in each choice set for our Bayesian D- and I-optimal designs for the different $\kappa$ values. For $\kappa = 0.5$, both types of designs score highly in utility balance, and there is hardly any difference between them in terms of utility balance. However, as $\kappa$ increases, the product of the probabilities drops, meaning that the designs become less utility balanced. 

\begin{figure}[ht]
    \centering
    \includegraphics[width=0.9\textwidth]{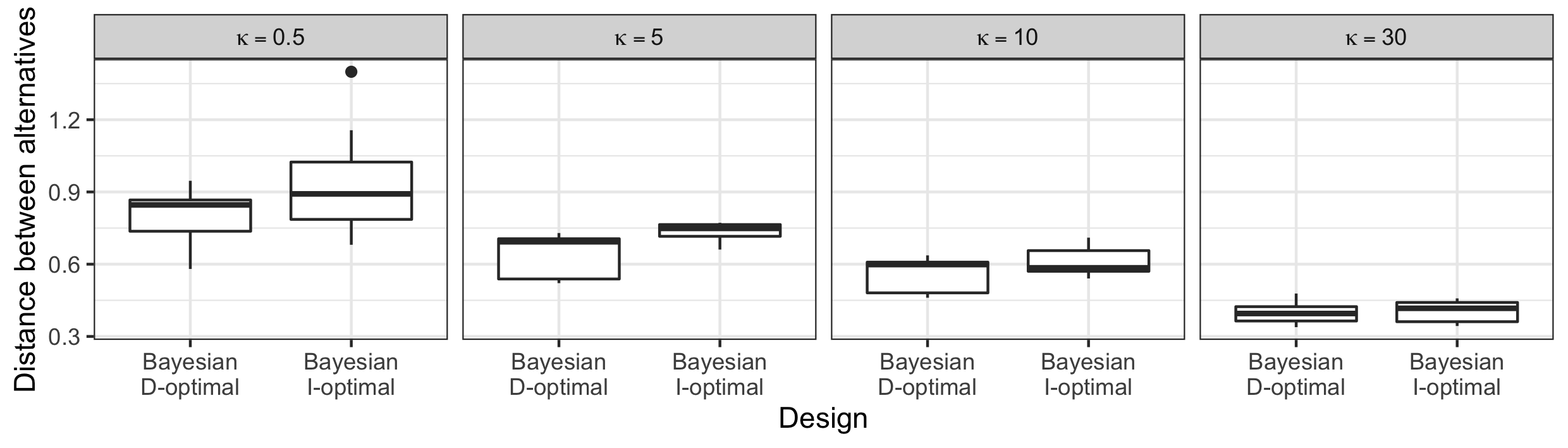}
    \caption{Euclidean distances between alternatives within a choice set for our Bayesian D- and I-optimal designs for the artificial sweetener experiment}
    \label{fig:res_cornell_analytic_transf_distances_within_choice_set}
\end{figure}

Figure \ref{fig:res_cornell_analytic_transf_distances_within_choice_set} shows boxplots of the Euclidean distances between the two alternatives within a choice set for our Bayesian D- and I-optimal designs. Just like in the cocktail experiment, mixtures within a single choice set tend to be closer together in the D-optimal designs than in the I-optimal designs, except when $\kappa=30$. For that value of $\kappa$, there is no major difference in the Euclidean distances between the alternatives within the choice sets of the D- and I-optimal designs. Finally, Figure \ref{fig:res_cornell_analytic_transf_distances_within_choice_set} shows that the distances between alternatives with choice sets decrease with the value of $\kappa$.

\section{Discussion}\label{sec:discussion}

In this paper, we introduced a computationally efficient definition for I-optimal designs for choice experiments and embedded the new I-optimality criterion in a coordinate-exchange algorithm for constructing I-optimal designs. By means of two examples from the literature, we demonstrated that the I-optimal designs perform substantially better than their D-optimal counterparts in terms of the variance of the predicted utility. We observed that I-optimal designs do not possess the utility balance property, which \citeauthor{huber1996importance} \cite{huber1996importance} considered to be desirable for efficient choice designs. However, \citeauthor{louviere2011design} \cite{louviere2011design} argued against it, suggesting to instead derive an optimal design using an appropriate prior distribution. We believe this to be a more sensible choice given that it is our interest to have designs that yield precise predictions for any combination of ingredient proportions.

While performing the research that led to this paper, we identified six possible extensions of our work. First of all, it would be of interest to extend the work presented here to other classes of models for data from mixture experiments than the Scheffé models, for example Becker models or Cox's mixture polynomial models \cite{cornell2002experiments}.

Second, the preference for a mixture may depend on characteristics other than its composition. For example, the ideal cocktail composition may also depend on the temperature at which it is served, or the most preferred bread might not only depend on the proportions of the various ingredients, but also on the baking time and the baking temperature. One practical example of such a scenario can be found in \citeauthor{zijlstra2019mixture} \cite{zijlstra2019mixture}, who observed that the preferred mobility budget mixture depends on the budget size. To cope with this kind of complication, our choice model for mixtures  must be extended to deal with the additional characteristics, typically called \textit{process variables} \cite{goos_jones_optimal_2011}.

Third, we focused on the multinomial logit model, which assumes that there is homogeneity in the preferences of the respondents. However, as demonstrated by \citeauthor{courcoux1997methode} \cite{courcoux1997methode} and \citeauthor{goos_hamidouche_2019_choice} \cite{goos_hamidouche_2019_choice}, this might be an unrealistic assumption. Therefore, it would make sense to extend the models and algorithms presented here to other types of models that take into account the possible presence of consumer heterogeneity. Examples of such models are the mixed logit model and the latent class choice model.

A fourth opportunity for future research is inspired by a practical difficulty that arises when conducting choice experiments with mixtures. When the number of distinct mixtures appearing in the Bayesian optimal designs is large and the mixtures have to be tasted, it is logistically very cumbersome to organize the experiment. For instance, organizing a choice experiment in which 40 distinct mixtures have to be tasted in perhaps 80 different choice sets is much harder to organize than a choice experiment in which only 20 distinct mixtures have to be tasted in 40 different choice sets. While the former experiment may be preferable from a statistical viewpoint, it may be practically infeasible. Developing an algorithm to find Bayesian I-optimal designs with mixtures with an upper bound on the number of distinct mixtures and/or an upper bound on the number of distinct choice sets is therefore valuable from a practitioner's point of view.

A fifth topic for future research would be to modify our coordinate-exchange algorithm, so that it can also cope with experimental regions that are not a simplex. Such experimental regions arise when there are constraints on the ingredient proportions other than lower bounds for individual proportions. Methodologically speaking, this is not highly innovative, since the mixture coordinate-exchange algorithm of \citeauthor{piepel_construction_2005} \cite{piepel_construction_2005} for linear regression models is already able to deal with this complication. However, embedding this capability in our implementation of the coordinate-exchange algorithm for choice experiments with mixtures would be useful for practitioners.

Finally, a sixth topic for future research would be to extend the I-optimality criterion to cope with model uncertainty. In this paper, we assumed that the model form is known at the planning stage of the experiment, which is not always the case in practice. Therefore, there is a risk of bias in the predictions due to model misspecification. A design and modeling strategy to deal with this possible misspecification would protect experimenters against such  bias.

\clearpage


\bibliographystyle{IEEEtranN}
\bibliography{bibliography}

\clearpage


\begin{appendices}

\section{Design tables of cocktail experiment}

In both tables, $x_1$, $x_2$, and $x_3$ denote the pseudocomponents that range from $0$ to $1$, while $a_1$, $a_2$, and $a_3$ are the ingredient proportions with their original lower bounds.

\begin{table}[ht]
\centering
\begin{tabular}{crrrrrr}
  \hline
Choice set & $x_1$ & $x_2$ & $x_3$ & $a_1$ & $a_2$ & $a_3$ \\ 
  \hline
1 & 0.00 & 1.00 & 0.00 & 0.30 & 0.60 & 0.10 \\ 
  1 & 0.60 & 0.40 & 0.00 & 0.57 & 0.33 & 0.10 \\ 
  2 & 1.00 & 0.00 & 0.00 & 0.75 & 0.15 & 0.10 \\ 
  2 & 0.46 & 0.00 & 0.54 & 0.51 & 0.15 & 0.34 \\ 
  3 & 0.00 & 0.00 & 1.00 & 0.30 & 0.15 & 0.55 \\ 
  3 & 0.00 & 0.55 & 0.45 & 0.30 & 0.40 & 0.30 \\ 
  4 & 0.00 & 0.00 & 1.00 & 0.30 & 0.15 & 0.55 \\ 
  4 & 0.00 & 0.55 & 0.45 & 0.30 & 0.40 & 0.30 \\ 
  5 & 0.60 & 0.40 & 0.00 & 0.57 & 0.33 & 0.10 \\ 
  5 & 0.00 & 1.00 & 0.00 & 0.30 & 0.60 & 0.10 \\ 
  6 & 1.00 & 0.00 & 0.00 & 0.75 & 0.15 & 0.10 \\ 
  6 & 0.40 & 0.60 & 0.00 & 0.48 & 0.42 & 0.10 \\ 
  7 & 0.00 & 1.00 & 0.00 & 0.30 & 0.60 & 0.10 \\ 
  7 & 0.00 & 0.41 & 0.59 & 0.30 & 0.34 & 0.36 \\ 
  8 & 1.00 & 0.00 & 0.00 & 0.75 & 0.15 & 0.10 \\ 
  8 & 0.31 & 0.36 & 0.33 & 0.44 & 0.31 & 0.25 \\ 
  9 & 0.00 & 0.50 & 0.50 & 0.30 & 0.38 & 0.32 \\ 
  9 & 0.40 & 0.29 & 0.31 & 0.48 & 0.28 & 0.24 \\ 
  10 & 0.00 & 1.00 & 0.00 & 0.30 & 0.60 & 0.10 \\ 
  10 & 0.36 & 0.33 & 0.31 & 0.46 & 0.30 & 0.24 \\ 
  11 & 0.00 & 0.50 & 0.50 & 0.30 & 0.38 & 0.32 \\ 
  11 & 0.40 & 0.29 & 0.31 & 0.48 & 0.28 & 0.24 \\ 
  12 & 0.40 & 0.60 & 0.00 & 0.48 & 0.42 & 0.10 \\ 
  12 & 1.00 & 0.00 & 0.00 & 0.75 & 0.15 & 0.10 \\ 
  13 & 0.49 & 0.00 & 0.51 & 0.52 & 0.15 & 0.33 \\ 
  13 & 0.00 & 0.48 & 0.52 & 0.30 & 0.37 & 0.33 \\ 
  14 & 0.52 & 0.00 & 0.48 & 0.53 & 0.15 & 0.32 \\ 
  14 & 0.27 & 0.40 & 0.33 & 0.42 & 0.33 & 0.25 \\ 
  15 & 0.00 & 0.00 & 1.00 & 0.30 & 0.15 & 0.55 \\ 
  15 & 0.53 & 0.00 & 0.47 & 0.54 & 0.15 & 0.31 \\ 
  16 & 0.54 & 0.46 & 0.00 & 0.54 & 0.36 & 0.10 \\ 
  16 & 0.30 & 0.33 & 0.38 & 0.43 & 0.30 & 0.27 \\ 
   \hline
\end{tabular}
\caption{Bayesian D-optimal design for the cocktail experiment} 
\label{tab:cocktail_exp_d_optimal_des}
\end{table}
\begin{table}[ht]
\centering
\begin{tabular}{crrrrrr}
  \hline
Choice set & $x_1$ & $x_2$ & $x_3$ & $a_1$ & $a_2$ & $a_3$ \\ 
  \hline
1 & 0.00 & 0.00 & 1.00 & 0.30 & 0.15 & 0.55 \\ 
  1 & 0.58 & 0.00 & 0.42 & 0.56 & 0.15 & 0.29 \\ 
  2 & 1.00 & 0.00 & 0.00 & 0.75 & 0.15 & 0.10 \\ 
  2 & 0.36 & 0.64 & 0.00 & 0.46 & 0.44 & 0.10 \\ 
  3 & 0.00 & 0.00 & 1.00 & 0.30 & 0.15 & 0.55 \\ 
  3 & 0.58 & 0.00 & 0.42 & 0.56 & 0.15 & 0.29 \\ 
  4 & 0.03 & 0.97 & 0.00 & 0.31 & 0.59 & 0.10 \\ 
  4 & 0.29 & 0.28 & 0.43 & 0.43 & 0.27 & 0.29 \\ 
  5 & 0.00 & 0.00 & 1.00 & 0.30 & 0.15 & 0.55 \\ 
  5 & 0.00 & 0.63 & 0.37 & 0.30 & 0.43 & 0.27 \\ 
  6 & 0.00 & 0.00 & 1.00 & 0.30 & 0.15 & 0.55 \\ 
  6 & 0.30 & 0.34 & 0.36 & 0.43 & 0.30 & 0.26 \\ 
  7 & 0.00 & 0.00 & 1.00 & 0.30 & 0.15 & 0.55 \\ 
  7 & 0.00 & 0.63 & 0.37 & 0.30 & 0.43 & 0.27 \\ 
  8 & 0.00 & 1.00 & 0.00 & 0.30 & 0.60 & 0.10 \\ 
  8 & 0.65 & 0.35 & 0.00 & 0.59 & 0.31 & 0.10 \\ 
  9 & 1.00 & 0.00 & 0.00 & 0.75 & 0.15 & 0.10 \\ 
  9 & 0.30 & 0.16 & 0.53 & 0.44 & 0.22 & 0.34 \\ 
  10 & 0.00 & 0.49 & 0.51 & 0.30 & 0.37 & 0.33 \\ 
  10 & 0.57 & 0.43 & 0.00 & 0.56 & 0.34 & 0.10 \\ 
  11 & 0.00 & 0.00 & 1.00 & 0.30 & 0.15 & 0.55 \\ 
  11 & 0.30 & 0.34 & 0.36 & 0.43 & 0.30 & 0.26 \\ 
  12 & 0.00 & 0.00 & 1.00 & 0.30 & 0.15 & 0.55 \\ 
  12 & 0.30 & 0.34 & 0.36 & 0.43 & 0.30 & 0.26 \\ 
  13 & 0.45 & 0.00 & 0.55 & 0.50 & 0.15 & 0.35 \\ 
  13 & 0.23 & 0.53 & 0.24 & 0.40 & 0.39 & 0.21 \\ 
  14 & 0.00 & 0.41 & 0.59 & 0.30 & 0.34 & 0.36 \\ 
  14 & 0.50 & 0.19 & 0.31 & 0.52 & 0.24 & 0.24 \\ 
  15 & 0.00 & 0.00 & 1.00 & 0.30 & 0.15 & 0.55 \\ 
  15 & 0.49 & 0.51 & 0.00 & 0.52 & 0.38 & 0.10 \\ 
  16 & 0.00 & 0.00 & 1.00 & 0.30 & 0.15 & 0.55 \\ 
  16 & 0.30 & 0.34 & 0.36 & 0.43 & 0.30 & 0.26 \\ 
   \hline
\end{tabular}
\caption{Bayesian I-optimal design for the cocktail experiment} 
\label{tab:cocktail_exp_i_optimal_des}
\end{table}

\clearpage

\section{Design tables of artificial sweetener experiment}

\begin{table}[ht]
\centering
\begin{tabular}{crrr}
  \hline
Choice set & $x_1$ & $x_2$ & $x_3$ \\ 
  \hline
1 & 0.00 & 1.00 & 0.00 \\ 
  1 & 0.61 & 0.39 & 0.00 \\ 
  2 & 0.00 & 1.00 & 0.00 \\ 
  2 & 0.24 & 0.30 & 0.47 \\ 
  3 & 0.00 & 0.00 & 1.00 \\ 
  3 & 0.00 & 0.67 & 0.33 \\ 
  4 & 1.00 & 0.00 & 0.00 \\ 
  4 & 0.40 & 0.60 & 0.00 \\ 
  5 & 0.22 & 0.44 & 0.33 \\ 
  5 & 0.59 & 0.00 & 0.41 \\ 
  6 & 0.40 & 0.00 & 0.60 \\ 
  6 & 1.00 & 0.00 & 0.00 \\ 
  7 & 0.00 & 0.33 & 0.67 \\ 
  7 & 0.48 & 0.25 & 0.27 \\ 
   \hline
\end{tabular}
\caption{Bayesian D-optimal design for the artificial sweetener experiment, when $\kappa = 0.5$} 
\label{tab:cornell_exp_d_optimal_des_kappa_0.5}
\end{table}
\begin{table}[ht]
\centering
\begin{tabular}{crrr}
  \hline
Choice set & $x_1$ & $x_2$ & $x_3$ \\ 
  \hline
1 & 0.00 & 0.00 & 1.00 \\ 
  1 & 0.39 & 0.34 & 0.27 \\ 
  2 & 0.00 & 1.00 & 0.00 \\ 
  2 & 0.02 & 0.00 & 0.98 \\ 
  3 & 1.00 & 0.00 & 0.00 \\ 
  3 & 0.10 & 0.70 & 0.21 \\ 
  4 & 0.00 & 0.00 & 1.00 \\ 
  4 & 0.52 & 0.00 & 0.48 \\ 
  5 & 0.00 & 0.00 & 1.00 \\ 
  5 & 0.39 & 0.34 & 0.27 \\ 
  6 & 0.52 & 0.48 & 0.00 \\ 
  6 & 0.17 & 0.15 & 0.68 \\ 
  7 & 0.06 & 0.00 & 0.94 \\ 
  7 & 0.00 & 0.51 & 0.49 \\ 
   \hline
\end{tabular}
\caption{Bayesian I-optimal design for the artificial sweetener experiment, when $\kappa = 0.5$} 
\label{tab:cornell_exp_i_optimal_des_kappa_0.5}
\end{table}
\begin{table}[ht]
\centering
\begin{tabular}{crrr}
  \hline
Choice set & $x_1$ & $x_2$ & $x_3$ \\ 
  \hline
1 & 0.00 & 0.00 & 1.00 \\ 
  1 & 0.28 & 0.28 & 0.43 \\ 
  2 & 0.56 & 0.44 & 0.00 \\ 
  2 & 0.28 & 0.30 & 0.42 \\ 
  3 & 0.00 & 1.00 & 0.00 \\ 
  3 & 0.00 & 0.48 & 0.52 \\ 
  4 & 1.00 & 0.00 & 0.00 \\ 
  4 & 0.51 & 0.49 & 0.00 \\ 
  5 & 0.00 & 0.68 & 0.32 \\ 
  5 & 0.42 & 0.32 & 0.26 \\ 
  6 & 1.00 & 0.00 & 0.00 \\ 
  6 & 0.50 & 0.00 & 0.50 \\ 
  7 & 0.56 & 0.00 & 0.44 \\ 
  7 & 0.26 & 0.41 & 0.32 \\ 
   \hline
\end{tabular}
\caption{Bayesian D-optimal design for the artificial sweetener experiment, when $\kappa = 5$} 
\label{tab:cornell_exp_d_optimal_des_kappa_5}
\end{table}
\begin{table}[ht]
\centering
\begin{tabular}{crrr}
  \hline
Choice set & $x_1$ & $x_2$ & $x_3$ \\ 
  \hline
1 & 0.00 & 0.00 & 1.00 \\ 
  1 & 0.47 & 0.00 & 0.53 \\ 
  2 & 0.27 & 0.43 & 0.31 \\ 
  2 & 0.08 & 0.00 & 0.92 \\ 
  3 & 1.00 & 0.00 & 0.00 \\ 
  3 & 0.39 & 0.36 & 0.25 \\ 
  4 & 0.42 & 0.27 & 0.31 \\ 
  4 & 0.00 & 0.09 & 0.91 \\ 
  5 & 0.48 & 0.52 & 0.00 \\ 
  5 & 0.18 & 0.19 & 0.62 \\ 
  6 & 0.00 & 1.00 & 0.00 \\ 
  6 & 0.33 & 0.42 & 0.24 \\ 
  7 & 0.00 & 0.00 & 1.00 \\ 
  7 & 0.00 & 0.51 & 0.49 \\ 
   \hline
\end{tabular}
\caption{Bayesian I-optimal design for the artificial sweetener experiment, when $\kappa = 5$} 
\label{tab:cornell_exp_i_optimal_des_kappa_5}
\end{table}
\begin{table}[ht]
\centering
\begin{tabular}{crrr}
  \hline
Choice set & $x_1$ & $x_2$ & $x_3$ \\ 
  \hline
1 & 0.00 & 0.00 & 1.00 \\ 
  1 & 0.25 & 0.24 & 0.51 \\ 
  2 & 0.00 & 1.00 & 0.00 \\ 
  2 & 0.00 & 0.57 & 0.43 \\ 
  3 & 0.55 & 0.00 & 0.45 \\ 
  3 & 1.00 & 0.00 & 0.00 \\ 
  4 & 0.54 & 0.00 & 0.46 \\ 
  4 & 0.28 & 0.38 & 0.34 \\ 
  5 & 0.57 & 0.43 & 0.00 \\ 
  5 & 1.00 & 0.00 & 0.00 \\ 
  6 & 0.55 & 0.45 & 0.00 \\ 
  6 & 0.31 & 0.33 & 0.37 \\ 
  7 & 0.00 & 0.68 & 0.32 \\ 
  7 & 0.37 & 0.36 & 0.27 \\ 
   \hline
\end{tabular}
\caption{Bayesian D-optimal design for the artificial sweetener experiment, when $\kappa = 10$} 
\label{tab:cornell_exp_d_optimal_des_kappa_10}
\end{table}
\begin{table}[ht]
\centering
\begin{tabular}{crrr}
  \hline
Choice set & $x_1$ & $x_2$ & $x_3$ \\ 
  \hline
1 & 0.00 & 0.00 & 1.00 \\ 
  1 & 0.39 & 0.01 & 0.60 \\ 
  2 & 0.51 & 0.25 & 0.24 \\ 
  2 & 0.98 & 0.00 & 0.02 \\ 
  3 & 0.00 & 0.10 & 0.90 \\ 
  3 & 0.38 & 0.26 & 0.36 \\ 
  4 & 0.20 & 0.28 & 0.53 \\ 
  4 & 0.44 & 0.55 & 0.01 \\ 
  5 & 0.27 & 0.41 & 0.32 \\ 
  5 & 0.12 & 0.00 & 0.88 \\ 
  6 & 0.00 & 0.00 & 1.00 \\ 
  6 & 0.01 & 0.41 & 0.58 \\ 
  7 & 0.00 & 0.96 & 0.04 \\ 
  7 & 0.27 & 0.52 & 0.21 \\ 
   \hline
\end{tabular}
\caption{Bayesian I-optimal design for the artificial sweetener experiment, when $\kappa = 10$} 
\label{tab:cornell_exp_i_optimal_des_kappa_10}
\end{table}
\begin{table}[ht]
\centering
\begin{tabular}{crrr}
  \hline
Choice set & $x_1$ & $x_2$ & $x_3$ \\ 
  \hline
1 & 1.00 & 0.00 & 0.00 \\ 
  1 & 0.68 & 0.12 & 0.20 \\ 
  2 & 0.00 & 0.00 & 1.00 \\ 
  2 & 0.16 & 0.17 & 0.67 \\ 
  3 & 0.51 & 0.00 & 0.49 \\ 
  3 & 0.37 & 0.28 & 0.35 \\ 
  4 & 0.00 & 1.00 & 0.00 \\ 
  4 & 0.00 & 0.66 & 0.34 \\ 
  5 & 0.00 & 0.51 & 0.49 \\ 
  5 & 0.29 & 0.33 & 0.38 \\ 
  6 & 0.47 & 0.53 & 0.00 \\ 
  6 & 0.41 & 0.31 & 0.28 \\ 
  7 & 0.31 & 0.69 & 0.00 \\ 
  7 & 0.00 & 1.00 & 0.00 \\ 
   \hline
\end{tabular}
\caption{Bayesian D-optimal design for the artificial sweetener experiment, when $\kappa = 30$} 
\label{tab:cornell_exp_d_optimal_des_kappa_30}
\end{table}
\begin{table}[ht]
\centering
\begin{tabular}{crrr}
  \hline
Choice set & $x_1$ & $x_2$ & $x_3$ \\ 
  \hline
1 & 0.10 & 0.59 & 0.31 \\ 
  1 & 0.00 & 0.32 & 0.68 \\ 
  2 & 0.38 & 0.00 & 0.62 \\ 
  2 & 0.66 & 0.08 & 0.26 \\ 
  3 & 0.24 & 0.54 & 0.22 \\ 
  3 & 0.00 & 0.81 & 0.19 \\ 
  4 & 0.80 & 0.00 & 0.20 \\ 
  4 & 0.56 & 0.25 & 0.19 \\ 
  5 & 0.06 & 0.26 & 0.68 \\ 
  5 & 0.00 & 0.00 & 1.00 \\ 
  6 & 0.00 & 0.00 & 1.00 \\ 
  6 & 0.26 & 0.07 & 0.67 \\ 
  7 & 0.44 & 0.50 & 0.06 \\ 
  7 & 0.29 & 0.36 & 0.35 \\ 
   \hline
\end{tabular}
\caption{Bayesian I-optimal design for the artificial sweetener experiment, when $\kappa = 30$} 
\label{tab:cornell_exp_i_optimal_des_kappa_30}
\end{table}
\end{appendices}

\end{document}